\documentclass[twocolumn,superscriptaddress,showpacs,aps,amsmath,amssymb]{revtex4}

\usepackage{amssymb}
\usepackage{mathrsfs}
\usepackage{epsfig}
\usepackage{amsbsy}
\usepackage{amsmath}
\usepackage{amsfonts}
\usepackage{graphicx}
\usepackage{latexsym}

\begin{document}
\title{Quantum Transport Theory for Photonic Networks}
\author{Chan U Lei}
\affiliation{Department of Physics and Center for Quantum
Information Science, National Cheng Kung University, Tainan 70101,
Taiwan }
\author{Wei-Min Zhang}
\email{wzhang@mail.ncku.edu.tw} \affiliation{Department of Physics
and Center for Quantum Information Science, National Cheng Kung
University, Tainan 70101, Taiwan }

\begin{abstract}
In this paper, we develop a quantum transport theory to describe
photonic transport in photonic networks. The photonic networks
concerned in the paper consist of all-optical circuits incorporating
photonic bandgap waveguides and driven resonators. The photonic
transport flowing through waveguides are entirely determined from
the exact master equation of the driven resonators. The master
equation of the driven resonators is obtained by explicitly
eliminating all the waveguide degrees of freedom while the
back-reactions between resonators and waveguides are fully taken
into account. The relations between the driven photonic dynamics and
photocurrents are obtained. The non-Markovian memory structure and
quantum coherence and decoherence effects in photonic transport are
also fully included. This quantum transport theory unifies two
fundamental nonequilibrium approaches, the Keldysh's nonequilibrium
Green function technique and the Feynman-Vernon influence functional
approach, together to make the investigation of the transient
quantum photonic transport become more powerful. As an illustration,
the theory is applied to the transport phenomena of a driven
nanocavity coupled to two waveguides in photonic crystals. The
controllability of photonic transport through the driven resonator
is demonstrated.
\end{abstract}

\date{March 1, 2011, revised}


\maketitle

\section{INTRODUCTION}

A photonic network is a communication network in which the
information is transmitted entirely in terms of optical signals
\cite{Opnetwork}. With the rapid development of nanotechnology, the
network in terms of all-optical circuits imbedded in photonic
crystals could be a promising device to provide a robust
interconnect network for optical communication
\cite{PC,manilightPC}. Photonic crystals are artificial materials
with periodic refractive index, its photonic band gap (PBG)
structure together with its characteristic dispersion properties
make the light manipulation and transmission much more efficient.
For examples, strong light confinement can be realized by
introducing point defect in photonic crystals \cite{HQcavity}, slow
light can be generated by waveguide structure with controllable
dispersion properties, which can be implemented by introducing line
defect or series of coupled point defects in photonic crystals
\cite{SlowLightObs,CROWproposal}. Various novel devices have been
constructed or proposed by the combination of different defects with
PBG structure, such as optical switches \cite{lowEswitch}, filters
\cite{WGfilter},
%low threshold lasers \cite{DefectLaser,LowThresholdLaser},
memory devices \cite{memory} and on-chip single photon gun
\cite{spg}, etc. Furthermore, with the new development of the
semiconductor nanofabrication technique, many tunable functional
photonic crystal devices have also recently been proposed and
modeled. Different techniques are utilized to characterize various
physical quantities in these photonic crystal devices. In
particular, properties such as the resonance frequency of a
resonator or the band structure of a waveguide can be tuned by
changing the refractive index of the photonic crystal through
thermooptic effect \cite{thermotunePBG}, electrooptical effect
\cite{electroopticcontrol}, fluid insertion \cite{intoptofluodic},
or even by mechanically changing the structure of photonic crystals
\cite{mectune}. Couplings between different elements in a photonic
circuit can be controlled with the help of
micro-/nanoelectromechanical systems \cite{nanomecchannaldrop}.
These dynamical tunable devices greatly expand the applications of
the photonic crystals for photonic integrated circuitry and further
stimulate the potential application in quantum information
processing.

To achieve the goal of quantum information processing with
all-optical processing, information carriers should be individual
photons and ultra-fast operations are necessary. Photonic
transmission processes in photonic networks should be able to vary
in an extremely quick and controllable way. In such a situation, all
the photonic devices are far away from equilibrium, where the
non-Markovian memory and quantum coherence and decoherence dynamics
dominate the photonic transport. Thus, a fundamental quantum
transport theory that can incorporate with the non-Markovian memory
and quantum coherence dynamics of photons for photonic transmission
in photonic networks is highly demanded.

In fact, non-Markovian dynamics in quantum optics has been
extensively studied for a few-level atom placed inside photonic
crystals \cite{Lam00455,Mog1089}. The typical features of the
non-Markovian dynamics include the atomic population trapping
(inhibition of spontaneous emission), the strong localization of
light, the formulation of atom-photon bound states and the
collective switching behavior in the vicinity of the PBG
\cite{Yab-Jon87,Joh902418,Kil92153,Kof94353,switching}. These
features can been found by solving exactly the Schr\"{o}dinger
equation for the atomic state contained only one photon or using a
perturbative expansion to the Heisenberg equation of motion in
powers of the atom-field reservoir coupling strength. When the
number of photons increases or the perturbation dos not work, the
problem becomes intractable and the general non-Markovian dynamics
with arbitrary number of photons at arbitrary temperature for the
structured reservoir has not been fully solved. Recently, we have
utilized the exact master equation of a micro/nano cavity coupled to
a general thermal reservoir and a structured reservoir in photonic
crystals to study non-perturbatively various non-Markovian process
involving arbitrary number of photons at arbitrary temperature
\cite{Xio10012105,Wu1018407}. However, a quantum photonic transport
theory in photonic networks consisting of all-optical circuits
incorporating photonic bandgap waveguides and driven resonators has
not been established in the literature.

A fundamental quantum transport theory should be built on a fully
nonequilibrium treatment. The modern nonequilibrium physics is
developed based on the Schwinger-Keldysh nonequilibrium Green
function technique \cite{Sch61407,Kad62} and the Feynman-Vernon
influence functional approach \cite{Fey63118}.  The nonequilibrium
Green function technique allows a systematic perturbative
\cite{Cho851,Rammer86323} and also a non-perturbative
\cite{zhang921900} study for various nonequilibrium phenomena in
many-body electronic systems, in particular in the steady limit. It
has become a very powerful tool in the study of steady quantum
electron transport in mesoscopic physics \cite{Hau98}. However, such
an approach has not been utilized to investigate photonic transport
in all-optical processings in photonic networks. Besides, the
problem of non-Markopvian memory structure and quantum decoherence
dynamics has not been well explored in terms of the
Schwinger-Keldysh nonequilibrium Green function technique in the
transient quantum transport.

On the other hand, the Feynman-Vernon influence functional approach
\cite{Fey63118} has been widely used to study dissipation dynamics
in quantum tunneling problems \cite{Leg871} and decoherence dynamics
in quantum measurement theory \cite{Zuk03715}. It is in particular
very useful to derive the exact master equation for the quantum
Brownian motion (QBM), achieved by integrating out completely the
environmental degrees of freedom through the path integral where the
non-Markovian memory structure and the decoherence processes are
manifested explicitly \cite{Cal83587,Hu922843}. The QBM is modeled
as a central harmonic oscillator linearly coupled to a set of
harmonic oscillators simulating a thermal bath. Applications of the
QBM exact master equation cover various topics, such as quantum
decoherence, quantum-to-classical transition, and quantum
measurement theory, etc. \cite{Wei99,Bre02,Hu08}. However, the
influence functional approach has not been fully used to formulate
in a closed form the problem of the quantum transport phenomena,
although some systematically expansion approaches, such as the
real-time diagrammatic expansion approach \cite{Sch9418436} and the
hierarchical equations-of-motion apprach \cite{Jin08234703}, have
been developed. Until very recently, we have established the exact
nonequilibrium theory for transient electron transport from the
exact master equation of nanodevices obtained based on the
Feynman-Vernon influence functional \cite{Jin10083013}.

In this paper, we shall attempt to develop a quantum transport
theory to depict photonic transport in photonic networks, based on
the recent development of the nonequilibrium quantum theory for the
transient electron transport dynamics in nanodevices
\cite{Jin10083013}. The photonic network consists of all-optical
circuits incorporating photonic bandgap waveguides and driven
resonators. We will focus on the dynamics of photonic transport
flowing from the resonators into waveguides through the controllable
photonic dynamics of the driven resonators. The photonic dynamics of
the driven resonators is determined by the exact master equation
where the waveguides are treated as reservoirs. The back-reactions
between the waveguides and the resonators and thereby the
non-Markovian dissipation and fluctuation induced from the
back-reactions are fully taken into account. We follow the similar
procedure of the nonperturbative derivation of the exact master
equation for nanoelectronics we developed recently
\cite{Tu08235311,Tu09631} to obtain the exact master equation for
photonic networks. The photocurrents flowing from the resonators
into the waveguides that describe the transient photonic transport
in the network are then obtained directly from the master equation.
The full response of the photonic transport under the control of
external driving fields is explicitly presented. The quantum kinetic
theory based on the Keldysh's nonequilibrium Green function
technique is reproduced and also generalized within our framework.

The remainder of the paper is organized as follow. In Sec.~II, we
model photonic networks in terms of a general open optical system,
and specify it by a photonic crystal structures in particular. The
fundamental Hamiltonian for the photonic network is derived from the
first-principle. The derivation of the exact master equation for the
driven resonators based on Feynman-Vernon influence functional
approach and the calculations of the relevant physical observables
from the master equation are presented in Sec.~III, where the
response of the resonator dynamics on the external driving field is
also explicitly given. In Sec.~IV, the photonic transport theory is
established. The photocurrents flowing from the resonators into
individual waveguides that describe the photonic transport in the
network are derived from the exact master equation, from which the
non-Markovian memory structure and the coherence and decoherence
dynamics in the photonic transport can be examined explicitly. A
comparison of the present theory with the electron transport in
mesoscopic systems based on the Keldysh's nonequilibrium Green
function technique is also given, from which the generalized lesser
(correlation) Green function which determines completely the quantum
kinetic theory of photonic networks is explicitly solved. In
Sec.~IV, we utilize the present theory to investigate quantum
transport phenomena of a driving nanocavity coupled to two
waveguides in photonic crystals, as an illustration. The
controllability of photonic transport through the driven resonator
is analytically and numerically demonstrated. Finally, summary and
prospective are given in Sec.~V. Some detailed derivations for the
master equation and analytical solutions in the weak coupling limit
are also given in Appendices.

\section{Modeling the photonic networks}

\begin{figure}
\includegraphics[width = 5.5 cm]{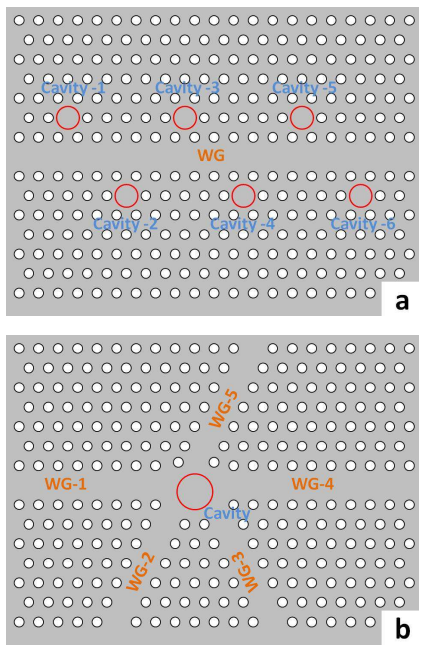}
\caption{Schematic plots of all-optical circuits incorporating
photonic bandgap waveguides and resonators. (a) Many nanocavities
couple to one waveguide at different sites; (b) One nanocavity (at
the center) couples to many individual waveguides.} \label{SysDem}
\end{figure}

The photonic network considered in this paper consists of photonic
band gap structure incorporating with defect cavities and waveguides
whose spectrums lie within the band gap of the photonic crystal, as
shown in Fig.~\ref{SysDem}. For the sake of simplicity, we restrict
ourselves to photonic crystal made up of linear, isotropic and
transparent medium. Therefore, the photonic network can be
characterized by a real scalar dielectric constant
$\epsilon(\mathbf{r})$ which is explicitly spatial dependent
\cite{PC}. To quantize the fields in the photonic network in terms
of the cavities and the waveguides modes, we adopt the
system-reservoir quantization formalism from \cite{FQ}, a
generalized canonical quantization of the electromagnetic field in
medium \cite{Gla91467}.
The electric and magnetic fields in the photonic network can be
expressed in terms of the vector potential
$\mathbf{A}(\mathbf{r},t)$ and a scalar potential
$\phi(\mathbf{r})$:
\begin{align}
& \mathbf{E}=-\nabla \phi - \frac{1}{c} \frac{
\partial\mathbf{A} }{\partial t} \ , ~~\mathbf{B}=\nabla \times \mathbf{A} \ .
\end{align}
We work in the Coulomb gauge and in the absence of source, which
corresponds to the choice of $\phi = 0$ with the transversality
condition $\nabla \cdot  [\epsilon(\mathbf{r})\mathbf{A}(\mathbf{r},
t)]=0$. Then the Hamiltonian of the electromagnetic field for the
photonic network is given by
\begin{equation}
\label{EMHam} H=\frac{1}{2}\int d^{3}\mathbf{r} \Big[\frac{c^2
\mathbf{\Pi}(\mathbf{r},t)^2}{\epsilon(\mathbf{r})} + (\nabla \times
\mathbf{A}(\mathbf{r},t))^2 \Big] \ ,
\end{equation}
where
$\mathbf{\Pi}(\mathbf{r},t)=\epsilon(\mathbf{r})\dot{\mathbf{A}}(\mathbf{r},t)
/c^2$ is the canonical conjugate of the vector potential.

We shall expand the vector potential in terms of a complete set of
mode functions which are defined by the wave equation
\begin{equation}
\nabla \times [\nabla \times \mathbf{f}_{k}(\mathbf{r})] -
\frac{\epsilon(\mathbf{r})\omega^2_k}{c^2}\mathbf{f}_{k}(\mathbf{r})
= \mathbf{0} \ ,
\end{equation}
where the mode functions $\mathbf{f}_{k}(\mathbf{r})$ satisfy the
orthonormality condition
\begin{equation}
\label{ScalarProduct} \int d^3\mathbf{r} \epsilon(\mathbf{r})
\mathbf{f}_{k}^*(\mathbf{r}) \cdot \mathbf{f}_{k'}(\mathbf{r}) =
\delta_{k,k'} \ .
\end{equation}
Utilizing the Feshbach projection technique \cite{FeshbachProj}, we
separate the whole space of the photonic network by the region
$\Im_{i}$ occupied by the cavity $i$,  the region $\Im_{\alpha}$
occupied by the waveguides $\alpha$, and the region left denoted by
$\Im_{b}$ as the background photonic crystal structure. Because of
the presence of the PBG structure, the micro/nano cavities in the
photonic crystal have very high $Q$ factor and the loss of the
waveguides in the photonic crystal is also extremely low, the field
in $\Im_{b}$ is then negligible as an effect of lossless materials
for photonic crystals. Consequently, the exact eigenmodes of the
vector potential in the photonic network can be decomposed into the
eigenmodes in the regions $\Im_{i}$ and $\Im_{\alpha}$, respectively
\begin{equation}
\mathbf{f}_{k}(\mathbf{r}) = \sum_{i}
\alpha_{i}^{k}\mathbf{u}_{i}(\mathbf{r}) + \sum_{\alpha} \sum_{k'}
\beta_{\alpha k'}^{k}\mathbf{v}_{\alpha k'}(\mathbf{r}) \ ,
\end{equation}
where $\mathbf{u}_{i}(\mathbf{r})$ is the eigenmode of cavity $i$,
$\mathbf{v}_{\alpha k}(\mathbf{r})$ is the the continuum eigenmode
of the waveguide $\alpha$. Both of them form a complete and
orthonormal basis set in regions $\Im_{i}$ and $\Im_{\alpha}$
separately. While $\alpha_{i}^{k}$ and $\beta_{\alpha}^{k}$ are the
expansion coefficients. The eigenmodes of the cavities and the
waveguides satisfy the following equations with the boundary
conditions \cite{FQ}
\begin{subequations}
\label{normm}
\begin{align}
 \nabla \times [\nabla & \times \mathbf{u}_{i}(\mathbf{r})] -
\frac{\epsilon(\mathbf{r})\omega_i^2}{c^2}\mathbf{u}_{i}(
\mathbf{r}) = \mathbf{0} \ , \\
& \mathbf{n} \times [\nabla \times \mathbf{u}_{i}(\mathbf{r})]
\mid_{\partial \Im_i} = \mathbf{0} \ ; \\
\nabla \times [\nabla & \times \mathbf{v}_{\alpha k}(\mathbf{r})] -
\frac{\epsilon(\mathbf{r})\omega_{\alpha k}^2}{c^2}
\mathbf{v}_{\alpha k}(\mathbf{r}) = \mathbf{0} \ , \\
& \mathbf{n} \times \mathbf{v}_{\alpha k}( \mathbf{r})\mid_{\partial
\Im_{\alpha}} = \mathbf{0} \ .
\end{align}
\end{subequations}

With the above decomposition of the exact eigenmodes, we can expand
the vector potential of the electromagnetic field in the photonic
network and its canonical conjugate in terms of the cavities
eigenmodes and the continuous modes of the waveguides
\begin{equation}
\begin{split}
& \mathbf{A}(\mathbf{r},t) = \sum_i \mathbf{A}_i (\mathbf{r},t)
+ \sum_\alpha \mathbf{A}_\alpha (\mathbf{r},t) \ , \\
& \mathbf{\Pi}(\mathbf{r},t) = \sum_i \mathbf{\Pi}_i (\mathbf{r},t)
+ \sum_\alpha \mathbf{\Pi}_\alpha (\mathbf{r},t) \ ,
\end{split}
\end{equation}
where $\mathbf{A}_i (\mathbf{r},t), \mathbf{\Pi}_i (\mathbf{r},t)$
and $\mathbf{A}_\alpha (\mathbf{r},t), \mathbf{\Pi}_\alpha
(\mathbf{r},t)$ are vector potential and canonical momentum of the
fields in cavity $i$ and waveguide $\alpha$, which can be quantized
following the standard quantum field theory in medium.  Without the
loss of generality and also for simplifying the notation, the
photonic network being considered consists of $N$ single mode
cavities and $M$ waveguides, each waveguide has a single continuous
spectrum (it is indeed straightforward to generalize the result to
the case of multi-mode cavities). Then we have
\begin{subequations}
\label{Expansion}
\begin{align}
 \mathbf{A}_i (\mathbf{r}, t)& =  c \sqrt{\frac{\hbar}{2
\omega_{i}}} ~\big[ a_{i} \mathbf{u}_{i}(\mathbf{r})e^{-i\omega_it}
+ a_{i}^\dag \mathbf{u}^{*}_{i}(\mathbf{r})e^{i\omega_it} \big] , \\
\mathbf{\Pi}_i (\mathbf{r}, t)& = -\frac{i\epsilon(\mathbf{r})}{c}
 \sqrt{\frac{\hbar\omega_i}{2}}~ \big[ a_{i}
\mathbf{u}_{i}(\mathbf{r}) e^{-i\omega_it}- a_{i}^\dag
\mathbf{u}^{*}_{i}(\mathbf{r})e^{i\omega_it} \big] , \\
\mathbf{A}_{\alpha} (\mathbf{r},t) & =  c \sum_k
\sqrt{\frac{\hbar}{2 \omega_k}} ~ \big[ c_{\alpha k}
\mathbf{v}_{\alpha k}(\mathbf{r})e^{-i\omega_k t}\notag \\
& ~~~~~~~~~~~~~~~~~~~ + c^{\dag}_{\alpha k}
\mathbf{v}^{*}_{\alpha k}(\mathbf{r}) e^{i\omega_k t}\big ] , \\
 \mathbf{\Pi}_{\alpha} (\mathbf{r},t) & =  -\frac{i\epsilon(\mathbf{r})}{c}
\sum_k \sqrt{\frac{\hbar\omega_k}{2}} ~ \big[ c_{\alpha k}
\mathbf{v}_{\alpha k}(\mathbf{r})e^{-i\omega_k t} \notag \\
& ~~~~~~~~~~~~~~~~~~~~~~~~~~~- c^{\dag}_{\alpha k}
\mathbf{v}^{*}_{\alpha k}(\mathbf{r})e^{i\omega_k t} \big] \ ,
\end{align}
\end{subequations}
where the operators $a_{i}, a_{i}^{\dag}$ and $c_{\alpha k},
c_{\alpha k}^{\dag}$ are the creation and annihilation operators of
the fields in cavity $i$ and waveguide $\alpha$. They obey the
following commutation relations
\begin{align}
[a_{i},a^{\dag}_{j}] & = \delta_{i j} \ , ~~[c_{\alpha k},
c^{\dag}_{\alpha' k'}] = \delta_{\alpha \alpha'}\delta_{k k'} \notag , \\
[a_{i},a_{j}]& =[c_{\alpha k},c_{\alpha' k'}] = [a_{i},c_{\alpha k}]
= [a^{\dag}_{i},c_{\alpha k}] = 0 .
\end{align}

Substitute the representations (\ref{Expansion}) into the
Hamiltonian (\ref{EMHam}), one arrives at the following Hamiltonian
for the photonic network
\begin{align}
H = & \sum_{i} \hbar \omega_{i} a^{\dag}_{i}a_{i} + \sum_{\alpha}
\sum_k \hbar \omega_k c^{\dag}_{\alpha k} c_{\alpha k} \notag \\ & +
\hbar \sum_{i} \sum_{\alpha, k} \big[ V_{i \alpha k} a^{\dag}_{i}
c_{\alpha k} + \widetilde{V}_{i \alpha k} a_{i} c_{\alpha k} + {\rm
H.c.} \big] \ ,
\end{align}
where $V_{i \alpha k}$ and $\widetilde{V}_{i \alpha k}$ are resonant
and nonresonant coupling constant between cavities and waveguides
which are defined by
\begin{equation}
\begin{split}
& V_{i \alpha k} = \frac{c^2}{2\sqrt{\omega_i \omega_k}}
\int_{\partial \Im_i \cap \partial \Im_\alpha}
[\mathbf{u}^{*}_i(\mathbf{r}) \times \mathbf{n} ] \cdot [ \nabla
\times \mathbf{v}_{\alpha k}(\mathbf{r})
] \ , \notag \\
& \tilde{V}_{i \alpha k} = \frac{c^2}{2\sqrt{\omega_i \omega_k}}
\int_{\partial \Im_i \cap \partial \Im_\alpha}
[\mathbf{u}_i(\mathbf{r}) \times \mathbf{n} ] \cdot [ \nabla \times
\mathbf{v}_{\alpha k}(\mathbf{r}) ] \ .
\end{split}
\end{equation}
Here we have ignored the couplings between cavities and also between
different waveguides because their spatial overlaps should be very
small. Also, due to the PBG structure in photonic crystals, the
features of high $Q$-factor micre/nano cavities and the lossless
waveguides allows us to neglect the nonresonant terms. Thus, the
Hamiltonian of the photonic network incorporating $M$ photonic
bandgap waveguides and $N$ resonators can be rewritten simple as
\begin{align}
 H = & \sum_{i=1}^N \hbar \omega_{i} a^{\dag}_{i}a_{i}
+ \sum_{\alpha=1}^M \sum_k \hbar \omega_{\alpha k} c^{\dag}_{\alpha
k} c_{\alpha k} \notag \\ & ~~~~ + \hbar \sum_{i} \sum_{\alpha, k}
\big( V_{i \alpha k} a^{\dag}_{i} c_{\alpha k} + H.c. \big) \ ,
\end{align}
The explicit form of the cavity and waveguide dispersions,
$\omega_i$ and $\omega_{\alpha k}$, and the coupling between the
cavities and waveguides, $V_{i\alpha k}$, depends on the detailed
implementation of the photonic network in photonic crystals, and can
be obtained by solving the eigenmode equation (\ref{normm}). It is
worth pointing out that the above derivation of the model
Hamiltonian can be extended straightforwardly to photonic networks
imbedded in other metamaterials.

Furthermore, to control the photonic transport in photonic networks,
we must incorporate the external driving fields applying to the
cavities and the waveguides, which can be realized by imbedding
light emitters such as quantum dot or nanowire to the cavities and
the waveguides \cite{WGfilter}. Then the Hamiltonian is modified as
\begin{subequations}
\label{HSEddf}
\begin{align}
H(t) = & \sum_{i} \hbar \omega_{i}a_{i}^{\dag }a_{i} + \sum_{i}
\big(f_{i}(t)a_{i}^{\dag} + f_{i}^{*}(t)a_{i}\big), \notag \\
& + \sum_{\alpha k}  \hbar\omega_{\alpha k}c_{\alpha k}^{\dagger
}c_{\alpha k} + \sum_{\alpha k} \big(f_{\alpha k}(t)c_{\alpha
k}^{\dag} + f_{\alpha
k}^{*}(t)c_{\alpha k}\big), \notag \\
& +  \hbar\sum_{i\alpha k} \big( V_{i \alpha k}(t)a_{i}^{\dag
}c_{\alpha k}+V_{i \alpha k}^*(t)c_{\alpha k}^{\dag }a_{i}\big).
\end{align}
\end{subequations}
This implementation is very similar to the electron transport in
mesoscopic systems in nanostructures \cite{Hau98,Jin10083013}. For
the electron transport in mesoscopic systems, one uses bias and gate
voltages applying to the electronic leads and the central region to
adjust the Fermi surfaces of the leads and the energy levels of the
central region. Here one has to use the external driving fields
directly acting to the cavities and the waveguides, which leads the
Hamiltonian to contain explicitly additional terms proportional to
the external driving fields.

Mathematically, we can transfer the external driving field
$f_{\alpha k}(t)$ acting on the waveguide $\alpha$ into an
equivalent external field acting on the cavities: $f_{i}(t)
\rightarrow \tilde{f}_{i}(t)=f_{i}(t)- V_{i \alpha k}(t)f_{\alpha
k}(t)/ \hbar\omega_{\alpha k}$ through a shift of the waveguide
operator $c_{\alpha k} \rightarrow \tilde{c}_{\alpha k}=c_{\alpha k}
+ f_{\alpha k}(t)/\hbar\omega_{\alpha k}$. Thus the general
Hamiltonian for photonic networks can be divided into three parts
\begin{align}
H(t) = H_S(t)+ \sum_\alpha H_{E\alpha}+ \sum_\alpha H_{T\alpha}(t) \
, \label{HM}
\end{align}
where,
\begin{subequations}
\label{Hmall}
\begin{align}
& H_{S}(t) =\sum_{i} \hbar \omega_{i}a_{i}^{\dag }a_{i} + \sum_{i}
\big(f_{i}(t)a_{i}^{\dag} + f_{i}^{*}(t)a_{i}\big) \ , \label{Hs} \\
& H_{E\alpha} = \sum_{k} \hbar \omega_{\alpha k}c_{\alpha
k}^{\dagger
}c_{\alpha k} \ , \\
& H_{T\alpha}(t) =  \hbar \sum_{ik} \big( V_{i \alpha
k}(t)a_{i}^{\dag }c_{\alpha k}+V_{i \alpha k}^*(t)c_{\alpha k}^{\dag
}a_{i}\big) \ .
\end{align}
\end{subequations}
The first part $H_{S}(t)$ is the Hamiltonian of the cavities in the
photonic network plus the contributions of the driving field,
$f_{i}(t)$, acting to the cavity $i$ (which also includes the
driving field $f_{\alpha k}(t)$ acting on the waveguides in terms of
the form of $- V_{i \alpha k}(t)f_{\alpha k}(t)/\omega_{\alpha k}$,
as shown above). The second part $H_{E}$ is the Hamiltonian of the
waveguides in the photonic network. The third part $H_{T}(t)$ is the
coupling between the cavities and the waveguides, which can be time
dependent in general through the controls of the coupling between
different elements, by means of dynamically tuning the photonic
structure \cite{manilightPC,nanomecchannaldrop}. With the help of
the general Hamiltonian (\ref{HM}), we can model various photonic
networks. For the sake of simplicity, we take $\hbar = 1$ hereafter.

\section{Exact master equation for driven resonators}

Experimentally, the photonic dynamics of  resonators and the
photonic transport between  waveguides and  resonators can be
controlled by the external driving field. When a resonator has
exchanges of particles, energy and information with the
surroundings, it becomes a typical open system. For an open quantum
system, its dynamics cannot be properly  described by
Schr\"{o}dinger's wave equation. It should be described by the
master equation of the reduced density matrix \cite{Car93}. The
reduced density matrix of the resonators, denoted by $\rho(t)$,
fully depicts the dynamics of photonic coherence of the driven
resonators coupled with many waveguides. Any other physical
observable is simply given by $\langle O(t)\rangle = \langle O
\rho(t)\rangle $.

The reduced density matrix describing the photonic quantum state of
the driven resonators is defined by tracing over all of the
waveguide degrees of freedom from the total system (the driven
resonators plus waveguides):
\begin{align}
\rho(t) ={\rm tr}_{E} [\rho_{\rm tot}(t)] \ ,
\end{align}
where $\rho_{\rm tot}(t)$ is the density matrix of the total system.
The total density matrix follows formally the evolution equation:
\begin{align}
\rho_{\rm tot}(t) = U(t,t_{0})\rho_{\rm tot}(t_{0})U^{\dag}(t,t_{0})
\ ,
\end{align}
where $U(t,t_{0})$ is the evolution operator,
\begin{align}
U(t,t_{0}) = T\exp \Big\{ -i\int_{t_{0}}^{t}H(\tau)d\tau \Big\},
\end{align} in which $T$ is the time-ordering operator, and
$H(\tau)$ is the Hamiltonian of the total system given by
Eq.~(\ref{HM}). In the following, we will derive the exact master
equation for the reduced density matrix by explicitly integrating
out all of the waveguide degrees of freedom based on the
Feynman-Vernon influence functional \cite{Fey63118}, similar to the
derivation of the master equation for electron coherence dynamics
and electron transport in nanostructures
\cite{Tu08235311,Tu09631,Jin10083013} and also for a coupled cavity
fields in a general bath \cite{An07042127}. The difference is that
here the system contains external driving fields and the reservoir
is at arbitrary finite temperature initially, which makes the
derivation much more complicated.

As usual, we assume that there is no initial correlation between the
resonators and waveguides \cite{Leg871}, namely $\rho_{tot}(t_{0}) =
\rho(t_{0}) \otimes \rho_{E}(t_{0})$, and the waveguides are
initially in the equilibrium state, i.e. $\rho_{E}(t_{0}) =
\frac{1}{Z}e^{- \sum_{\alpha} \beta_{\alpha}H_{E \alpha}}$, with
$\beta_{\alpha} = 1/(k_{B}T_{\alpha})$ and $T_{\alpha}$ is the
initial temperature of the waveguide $\alpha$. Then the reduced
density matrix at arbitrary later time $t$ can be expressed in the
coherent state representation as \cite{An07042127,Tu08235311}:
\begin{align}
\label{coherho} & \langle \boldsymbol{\alpha}_{f}  |
\rho(t)| \boldsymbol{\alpha}'_{f} \rangle \notag \\
& = \int
d\mu(\boldsymbol{\alpha}_{0})d\mu({\boldsymbol{\alpha}'_{0}})
\langle \boldsymbol{\alpha}_{0} | \rho(t_{0}) |
\boldsymbol{\alpha}'_{0} \rangle \mathcal{J}(
\boldsymbol{\alpha}_{f}^{*} , \boldsymbol{\alpha}'_{f}, t |
\boldsymbol{\alpha}_{0}, {\boldsymbol{\alpha}'}_{0}^{*}, t_{0} )
\end{align}
with the vector $\boldsymbol{\alpha} = (\alpha_{1}, \alpha_{2},
...)$ and $|\boldsymbol{\alpha}\rangle = |\alpha_{1}\rangle
|\alpha_{2}\rangle...$ is a unnormalized multi-mode coherent states,
i.e. $a_{i} |\boldsymbol{\alpha} \rangle = \alpha_{i}
|\boldsymbol{\alpha} \rangle$ and $\langle
\boldsymbol{\alpha}|\boldsymbol{\alpha}'\rangle =
\exp(\sum_i\alpha^*_i\alpha'_i) = \exp(\boldsymbol{\alpha}^\dag
\boldsymbol{\alpha}')$, while
$d\mu(\boldsymbol{\alpha})=\frac{d\boldsymbol{\alpha}^*
d\boldsymbol{\alpha}}{2 \pi i}e^{-|\boldsymbol{\alpha}|^2}$ is the
integral measure of the Bergmann complex space \cite{zhang90867}.
The propagating function of the reduced density matrix in
Eq.~(\ref{coherho}) is given in terms of path integrals:
\begin{align}
\label{propagatingfunc} & \mathcal{J}( \boldsymbol{\alpha}_{f}^{*} ,
\boldsymbol{\alpha}'_{f}, t | \boldsymbol{\alpha}_{0},
{\boldsymbol{\alpha}'}_{0}^{*}, t_{0} ) \notag \\ & = \int {\cal
D}[\boldsymbol{\alpha}^{*} \boldsymbol{\alpha} ;
{\boldsymbol{\alpha}'}^{*} \boldsymbol{\alpha}'] e^{i(
S_{S}[\boldsymbol{\alpha}^{*}, \boldsymbol{\alpha}] -
S_{S}[{\boldsymbol{\alpha}'}^{*}, \boldsymbol{\alpha}'] )}
\mathscr{F}[\boldsymbol{\alpha}^{*} \boldsymbol{\alpha} ;
{\boldsymbol{\alpha}'}^{*} \boldsymbol{\alpha}']
\end{align}
with the integral boundary conditions:
$\boldsymbol{\alpha}(t_0)=\boldsymbol{\alpha}_{0}$,
$\boldsymbol{\alpha}^{*}(t)=\boldsymbol{\alpha}_{f}^{*}$,
${\boldsymbol{\alpha}'}^{*}(t_0)={\boldsymbol{\alpha}'}_{0}^{*}$,
and $\boldsymbol{\alpha}'(t)=\boldsymbol{\alpha}'_{f}$, where
$S_{S}[\boldsymbol{\alpha}^{*}, \boldsymbol{\alpha}]$ is the action
of the driven resonators in the coherent state representation:
%\begin{subequations}
\begin{align}
S_{S}[\boldsymbol{\alpha}^{*},  \boldsymbol{\alpha}] = & -
\frac{i}{2}[\boldsymbol{\alpha}^{\dag}_f \boldsymbol{\alpha}(t) +
\boldsymbol{\alpha}^{\dag}(t_{0}) \boldsymbol{\alpha}_0] +
\int_{t_{0}}^{t} d\tau \Big\{
\frac{i}{2}\Big[\boldsymbol{\alpha}^{\dag} \frac{d
\boldsymbol{\alpha}}{d\tau} \notag \\ & - \frac{d
\boldsymbol{\alpha}^{\dag}}{d\tau} \boldsymbol{\alpha}\Big]-
(\boldsymbol{\alpha}^{\dag} \boldsymbol{\omega}\boldsymbol{\alpha}+
\boldsymbol{\alpha}^{\dag} \boldsymbol{f} + \boldsymbol{f}^{\dag}
\boldsymbol{\alpha} )\Big\} ,
\end{align}
in which the frequency matrix $\boldsymbol{\omega} \equiv
\{\omega_{ij} \}$ (here the off-diagonal matrix elements
representing the possible couplings between different cavities are
included for the generality of the formulation) and the driving
field vector $\boldsymbol{f}(\tau) \equiv \{ f_i(\tau) \}$. While
$\mathscr{F}[\boldsymbol{\alpha}^{*} \boldsymbol{\alpha} ;
\boldsymbol{\alpha}'^{*} \boldsymbol{\alpha}']$ is the influence
functional \cite{Fey63118} obtained after integrated out all of the
waveguide degrees of freedom \cite{An07042127,Tu08235311}:
\begin{widetext}
\begin{align}
\label{influencefunctional}  \mathscr{F}[\boldsymbol{\alpha}^{*} &
\boldsymbol{\alpha} ; {\boldsymbol{\alpha}'}^{*}
\boldsymbol{\alpha}']
 = \exp \Big\{ \sum_{\alpha} ( - \int_{t_{0}}^{t}d\tau
\int_{t_{0}}^{\tau} d\tau' \boldsymbol{\alpha}^{\dag}(\tau)
\mathbf{g}_{\alpha}(\tau, \tau') \boldsymbol{\alpha}(\tau')  -
\int_{t_{0}}^{t}d\tau \int_{t_{0}}^{\tau} d\tau'
\boldsymbol{\alpha}'^{\dag}(\tau') \mathbf{g}_{\alpha}(\tau', \tau)
\boldsymbol{\alpha}'(\tau) \notag \\ & + \int_{t_{0}}^{t}d\tau
\int_{t_{0}}^{t} d\tau' {\boldsymbol{\alpha}'}^{\dag}(\tau)
\mathbf{g}_{\alpha}(\tau, \tau') \boldsymbol{\alpha}(\tau')  -
\int_{t_{0}}^{t}d\tau \int_{t_{0}}^{t} d\tau' [
\boldsymbol{\alpha}^{\dag}(\tau) -
{\boldsymbol{\alpha}'}^{\dag}(\tau)
]\widetilde{\mathbf{g}}_{\alpha}(\tau, \tau') [
\boldsymbol{\alpha}(\tau') - \boldsymbol{\alpha}'(\tau') ] ) \Big\}
.
\end{align}
%\end{subequations}
\end{widetext}
The time-correlation functions in Eq.~(\ref{influencefunctional})
are given by:
\begin{subequations}
\label{t-correlations}
\begin{align}
\mathbf{g}_{\alpha ij}(\tau, \tau') & = \sum_{k} V_{i \alpha
k}(\tau)V_{j \alpha k}^{*}(\tau')e^{-i \omega_{\alpha
k}(\tau-\tau')} \ , \\
\widetilde{\mathbf{g}}_{\alpha ij}(\tau, \tau')& = \sum_{k} V_{i
\alpha k}(\tau)V_{j \alpha k}^{*}(\tau') n_{\alpha}(\omega_{\alpha
k}) e^{-i\omega_{\alpha k}(\tau-\tau')} ,
\end{align}
\end{subequations}
which depict the time correlations of photons in the waveguides
through the resonators, and $n_{\alpha}(\omega_{\alpha k}) = 1/(
e^{\beta_{\alpha} \omega_{\alpha k}} - 1 )$ is the initial thermal
photonic distribution function in the waveguide $\alpha$ at the time
$t_{0}$. The influence functional, Eq.~(\ref{influencefunctional}),
contains all the back reactions from waveguides to the driven
resonators through the photonic transfer between the resonators and
waveguides. If the coupling constants do not explicitly depend on
the time, then introducing the spectral density
$J_\alpha(\omega)=2\pi\sum_{k} V_{i \alpha k}V_{j \alpha k}^{*}
\delta(\omega-\omega_{\alpha k})$, we can simply express the
time-correlation function in terms of the spectral density as
follows
\begin{subequations}
\label{spectralf}
\begin{align}
\mathbf{g}_{\alpha ij}(\tau, \tau') & = \int \frac{d\omega}{2\pi}
J_\alpha(\omega)e^{-i \omega(\tau-\tau')} \ ,\\
\widetilde{\mathbf{g}}_{\alpha ij}(\tau, \tau')& = \int
\frac{d\omega}{2\pi} J_\alpha(\omega) n_{\alpha}(\omega)
e^{-i\omega(\tau-\tau')} \ .
\end{align}
\end{subequations}

As shown in Eqs.~(\ref{influencefunctional}) and
(\ref{propagatingfunc}), after integrated out all of the waveguide
degrees of freedom, the effective action in the propagating function
remains in a quadratic form. Therefore, the path integral in the
propagating function can be carried out exactly through the
stationary path method. The resulting propagating function,
Eq.~(\ref{propagatingfunc}), becomes (see a detailed derivation
given in Appendix A)
\begin{align}
\label{propagatingfunc2} \mathcal{J}( & \boldsymbol{\alpha}^{*}_{f}
, \boldsymbol{\alpha}'_{f}, t | \boldsymbol{\alpha}_{0},
{\boldsymbol{\alpha}'}_{0}^{*}, t_{0} ) =  A(t)\exp\{(
\boldsymbol{\alpha}^{\dag}_{f}-
\boldsymbol{y}^{\dag}(t))\boldsymbol{J}_{1}(t)
\boldsymbol{\alpha}_{0} \notag \\
& + {\boldsymbol{\alpha}'}_{0}^{\dag} \boldsymbol{J}_{1}^{\dag}(t)
(\boldsymbol{\alpha}'_{f}-\boldsymbol{y}(t))  +
\boldsymbol{\alpha}^{\dag}_{f} \boldsymbol{J}_{2}(t)
\boldsymbol{\alpha}'_{f}  + {\boldsymbol{\alpha}'}^{\dag}_{0}
\boldsymbol{J}_{3}(t) \boldsymbol{\alpha}_{0} \notag \\ & -
\boldsymbol{\alpha}^{\dag}_{f} \boldsymbol{y}(t)
-\boldsymbol{y}^{\dag}(t) \boldsymbol{\alpha}'_{f} -
\boldsymbol{y}^{\dag}(t) \boldsymbol{w}(t) \boldsymbol{y}(t) \},
\end{align}
in which $A(t)=\det[\boldsymbol{w}(t)]$, $\boldsymbol{J}_{1}(t) =
\boldsymbol{w}(t)\boldsymbol{u}(t,t_{0})$, $\boldsymbol{J}_{2}(t) =
1 - \boldsymbol{w}(t)$, $\boldsymbol{J}_{3}(t) = 1 -
\boldsymbol{u}^{\dag}(t,t_{0})\boldsymbol{w}(t)\boldsymbol{u}(t,t_{0})$
and $\boldsymbol{w}(t) = [\boldsymbol{1} + \boldsymbol{v}(t,t)]^{-1}
= \boldsymbol{w}^{\dag}(t)$. The function $\boldsymbol{u}(\tau,
t_{0})$, $\bar{\boldsymbol{u}}(\tau,t)$, $\boldsymbol{v}(\tau, t)$
are $N$ by $N$ matrices, and $\boldsymbol{y}(t)$ is a $1 \times N$
matrix, where $N$ is the number of resonator modes. These functions
obey the equations of motion
\begin{subequations}
\label{sol2}
\begin{align}
& \label{sol2b} \frac{d\boldsymbol{u}(\tau, t_{0})}{d\tau} +
i\boldsymbol{\omega}\boldsymbol{u}(\tau, t_{0}) +
\int_{t_{0}}^{\tau} \mathbf{g}(\tau, \tau')\boldsymbol{u}(\tau',
t_{0})d\tau' = 0 \ ,
\\ & \label{sol2a} \frac{d\bar{\boldsymbol{u}}(\tau, t)}{d\tau} +
i\boldsymbol{\omega}\bar{\boldsymbol{u}}(\tau, t) - \int_{\tau}^{t}
\mathbf{g}(\tau, \tau')\bar{\boldsymbol{u}}(\tau', t)d\tau' = 0 \ ,
\\ &
\label{sol2c} \frac{d\boldsymbol{v}(\tau, t)}{d\tau} +
i\boldsymbol{\omega}\boldsymbol{v}(\tau, t) + \int_{t_{0}}^{\tau}
\mathbf{g}(\tau, \tau')\boldsymbol{v}(\tau', t)d\tau' \notag \\ &
~~~~~~~~~~~~~~~~~~~~~~ ~~~~~~~~ = \int_{t_{0}}^{t}
\widetilde{\mathbf{g}}(\tau, \tau')\bar{\boldsymbol{u}}(\tau', t)
d\tau' \ ,
\\ &
\label{sol2d} \frac{d\boldsymbol{y}(\tau)}{d\tau} +
i\boldsymbol{\omega}\boldsymbol{y}(\tau) + \int_{t_{0}}^{\tau}
\mathbf{g}(\tau, \tau')\boldsymbol{y}(\tau')d\tau' =
-i\boldsymbol{f}(\tau)
\end{align}
\end{subequations}
subjected to the initial conditions $\bar{\boldsymbol{u}}(t, t) =
1$, $\boldsymbol{u}(t_{0}, t_{0}) = 1$,  $\boldsymbol{v}(t_{0}, t) =
0$ and $\boldsymbol{y}(t_{0}) = 0$ with $t_0\leq \tau,\tau'\leq t$.
Here we have also defined $\mathbf{g}(\tau, \tau')=\sum_\alpha
\mathbf{g}_{\alpha}(\tau, \tau')$ and $\widetilde{\mathbf{g}}(\tau,
\tau')=\sum_\alpha \widetilde{\mathbf{g}}_{\alpha}(\tau, \tau')$. It
is not too difficult to find that
\begin{subequations}
\label{sbuyv}
\begin{align}
& \bar{\boldsymbol{u}}(\tau, t) = \boldsymbol{u}^{\dag}(t, \tau) \ , \label{ubars}\\
& \boldsymbol{y}(\tau) = -i \int_{t_{0}}^{\tau} \boldsymbol{u}(\tau,
\tau') \boldsymbol{f}(\tau') d\tau' \ , \label{difield}
\\ & \boldsymbol{v}(\tau,t) = \int_{t_{0}}^{\tau}d\tau'
\int_{t_{0}}^{t}d\tau'' \boldsymbol{u}(\tau, \tau')
\widetilde{\mathbf{g}}(\tau', \tau'')\bar{\boldsymbol{u}}(\tau'', t)
\ . \label{vtt}
\end{align}
\end{subequations}
As we see, once Eq.~(\ref{sol2b}) is solved for
$\boldsymbol{u}(\tau, t_{0})$, the dynamics of the driven resonators
can be completely determined.

Substituting Eq.~(\ref{propagatingfunc2}) into Eq.~(\ref{coherho}),
taking a time derivative to the reduced density matrix, and then
using the relations \cite{zhang90867}:
$a_{i}|\boldsymbol{\alpha}\rangle =
\alpha_{i}|\boldsymbol{\alpha}\rangle$ and $\frac{\partial}{\partial
\alpha_{i}}|\boldsymbol{\alpha}\rangle =
a_{i}^{\dag}|\boldsymbol{\alpha}\rangle$, we obtain a time
convolutionless but exact master equation for the driving resonator
system coupled to waveguides:
\begin{align}
\label{ME} & \frac{d\rho(t)}{dt} = -i [H_{\rm eff}(t), \rho(t)]
\notag
\\ & + \sum_{ij}\gamma_{ij}(t)[
2a_{j} \rho(t)a_{i}^{\dag} - \rho(t) a_{i}^{\dag}a_{j} -
a_{i}^{\dag}a_{j}\rho(t) ] \notag
\\ & + \sum_{ij} \widetilde{\gamma}_{ij}(t) [
a_{j}\rho(t)a_{i}^{\dag} + a_{i}^{\dag}\rho(t)a_{j} -
a_{i}^{\dag}a_{j}\rho(t) - \rho(t)a_{j}a_{i}^{\dag} ] ,
\end{align}
where $H_{\rm eff}(t)=  \sum_{ij} \omega_{ij}'(t) a^{\dag}_{i} a_{j}
+ \sum_{i}[f'_{i}(t)a^{\dag}_{i} + {f'}_{i}^{*}(t)a_{i}]$ is the
effective Hamiltonian of the driven resonators. The renormalized
frequencies $\omega_{ij}'(t)$ and the renormalized driving field
 $f'_{i}(t)$ result from the back-reaction effect of the
waveguides coupled to the resonators. The initial temperature
effects are fully incorporated into the temperature-dependent noise
coefficient $\widetilde{\boldsymbol{\gamma}}(t)$. The time dependent
dissipation coefficients $\boldsymbol{\gamma}$(t) together with the
time-dependent noise coefficients
$\widetilde{\boldsymbol{\gamma}}(t)$ describe the non-Markovian
dissipation and decoherence dynamics of the resonators due to the
interaction between the resonators and waveguides. Here these
coefficients also describe the photon transport from the resonators
into the waveguides under the control of external driving fields, as
we will show in the next section. All these time-dependent
coefficients in Eq.~(\ref{ME}) are given explicitly as follow:
\begin{subequations}
\label{coeff}
\begin{align}
\omega'_{ij}(t) &  = \omega_{ij}(t) - \frac{i}{2}
\sum_{\alpha}[\boldsymbol{\kappa}_{\alpha}(t) -
\boldsymbol{\kappa}_{\alpha}^{\dag}(t)]_{ij} \ , \\
\gamma_{ij}(t) & = \frac{1}{2}
\sum_{\alpha}[\boldsymbol{\kappa}_{\alpha}(t) +
\boldsymbol{\kappa}_{\alpha}^{\dag}(t)]_{ij}\ , \\
\widetilde{\gamma}_{ij}(t) & =
\sum_{\alpha}[\boldsymbol{\lambda}_{\alpha}(t) +
\boldsymbol{\lambda}_{\alpha}^{\dag}(t)]_{ij} \ , \\
f'_i(t) & =f_i(t) + \sum_{\alpha} [\boldsymbol{f}_{\alpha}(t)]_i
\end{align}
\end{subequations}
with
\begin{subequations}
\label{coeff2}
\begin{align}
\boldsymbol{\kappa}_{\alpha} (t) & = \int_{t_{0}}^{t} d\tau
\mathbf{g}_{\alpha}(t, \tau) \boldsymbol{u}(\tau, t_{0})
\boldsymbol{u}^{-1}(t, t_{0}) \ ,
\\
\boldsymbol{\lambda}_{\alpha}(t) & =
\int_{t_{0}}^{t}d\tau[\mathbf{g}_{\alpha}(t,\tau)\boldsymbol{v}(\tau,t)
- \widetilde{\mathbf{g}}_{\alpha}(t, \tau)\bar{\boldsymbol{u}}(\tau,
t)] \notag \\
&~~~~~~~~~~~~~~~- \boldsymbol{\kappa}_{\alpha}(t)\boldsymbol{v}(t,t)
\ , \\
\boldsymbol{f}_{\alpha}(t) & = i
\boldsymbol{\kappa}_{\alpha}(t)\boldsymbol{y}(t) - i
\int_{t_0}^{t}d\tau\mathbf{g}_{\alpha}(t,\tau)\boldsymbol{y}(\tau)
\end{align}
\end{subequations}
as functions of $\boldsymbol{u}(t,t_{0})$, $\boldsymbol{v}(t,t)$ and
$\boldsymbol{y}(t)$ that are determined by the integrodifferential
equations of Eq.~(\ref{sol2}). We should point out that although it
has a  time convolutionless form, the master equation is exact and
the non-Markovian memory effect between the resonators and
waveguides is fully embodied non-perturbatively in the integral
kernels involving the non-local time-correlation functions of the
waveguides, $\mathbf{g}(\tau, \tau')$ and
$\widetilde{\mathbf{g}}(\tau, \tau')$, in Eq.~(\ref{sol2}). We
should also point out that $\boldsymbol{f}_{\alpha}(t)$ is a shift
to the external driving field $\boldsymbol{f}(t)$, which comes from
the back-reaction of the waveguide $\alpha$ to the resonators. In
other words, $\boldsymbol{f}_{\alpha}(t)$ is a feed-back effect to
the external driving field. We may call the function
$\boldsymbol{y}(t)$ (see Eq.~(\ref{difield})) as the driving-induced
field.

Notice that using the Feynman-Vernon influence functional approach
to derive the exact master equation was carried out early for
quantum Brown motion (QBM) \cite{Hu922843,Kar97153}. The exact
master equation can also be obtained by the trace-over-bath on
total-space Wigner-function method
\cite{Haa852462,Hal962012,For01105020} or the stochastic diffusion
Schr\"{o}dinger equation \cite{Str04052115}. All these previous
works deal mainly with a single harmonic oscillator coupled to a
thermal bath without external driving fields. The extension of the
exact master equation to the systems of two entangled optical modes
or two entangled harmonic oscillators has only been worked out very
recently \cite{An07042127,Cho08011112,Paz08220401}. The exact master
equation of Eq.~(\ref{ME}) obtained here is indeed the most general
one for the system containing arbitrary number of entangled modes
coupled to arbitrary number of photonic reservoirs with arbitrary
spectral density at arbitrary initial temperatures under arbitrary
number of external driving fields. It allows to investigate various
general and exact non-Markovian dynamics in photonic systems.

On the other hands, one may obtain a time-convolutionless master
equation from perturbation expansion up to the second order in terms
of the coupling constant between the system and the reservoir
\cite{Bre02,Car93}. Such a time-convolutionless master equation is
valid only in the weak coupling regime where the non-Markovian
effect is also very weak, except for some structured reservoirs
\cite{switching}. Also, the time-convolutionless perturbation master
equation up to the second order in the system-reservoir coupling is
indeed the same as the Born-Markov master equation without taking
the long time limit \cite{Car93,Xio10012105}. An exception of the
time convolutionless exact master equation not for pure optical
systems is the master equation of a two-level atom with one single
photon at zero temperature \cite{Bre991633,An10052330}. However,
when the number of photons increases, the problem becomes
intractable and the general non-Markovian dynamics with arbitrary
number of photons at arbitrary temperature has not been well
explored in quantum optics. Different from the study on the
non-Markovian atomic dynamics involving only one single photon in
quantum optics, the exact master equation of Eq.~(\ref{ME}) is
capable to investigate the general and the exact non-Markovian
process involving arbitrary number of photons of the reservoir with
arbitrary spectral density at arbitrary temperature in all-optical
circuits. Applications to such exact non-Markovian dynamics of a
micro/nano cavity coupled to a general thermal reservoir and a
structured reservoir in photonic crystals have just
 been worked out very recently \cite{Xio10012105,Wu1018407}.

In fact, the master equation (\ref{ME}) determines all the
nonequilibrium dynamics of the driven resonators and the photonic
transport between the resonators and the waveguides in photonic
networks. The aim of the present paper is to use the exact master
equation derived above to establish a quantum transport theory that
generalizes the quantum transport theory based on the Keldysh's
nonequilibrium Green function technique. We will show in the next
section how the photocurrent passing through the resonators into
waveguides can be obtained directly from the master equation, and
how the quantum transport theory based on the Keldysh's
nonequilibrium Green function technique can be easily reproduced and
generalized. Before end this section, we shall calculate some
important physical observables describing the photonic dynamics in
the driven resonators. One of them is the time evolution of the
photonic resonator field, and another one is the single particle
reduced density matrix which characterizes photonic intensity of the
resonators. These are two main quantities related to the
non-Markovian transport dynamics in photonic systems.

The photonic field of the resonators is determined by $\langle
a_{i}(t) \rangle = {\rm tr}_s[ a_{i} \rho (t) ]$. With the help of
the master equation, it obeys the equation of motion:
\begin{equation}
\label{aEOM} \frac{d}{dt} \langle a_{i}(t) \rangle = \sum_{j}
[\dot{\boldsymbol{u}}(t,t_{0})\boldsymbol{u}^{-1}(t,t_{0})]_{ij}
\langle a_{j}(t) \rangle - if'_{i}(t) \ .
\end{equation}
Its solution is:
\begin{equation}
\label{aEOMsol} \langle a_{i}(t) \rangle = \sum_{j}
u_{ij}(t,t_{0})\langle a_{j}(t_{0}) \rangle + y_{i}(t) \ .
\end{equation}
In other words, the photonic field of the driven resonators is a
combination of the photonic propagating  of the initial field
$\boldsymbol{u}(t,t_{0})\langle a_{j}(t_{0}) \rangle$ and the
driving-induced field $\boldsymbol{y}(t)$ (see Eq.~(\ref{difield})).
This solution describes the driving-field-induced photonic coherence
in resonators. Similarly, the equation of motion for the single
particle reduced density matrix of the resonators,
$\rho^{(1)}_{ij}(t) = {\rm tr}_s[ a^{\dag}_{j} a_{i}\rho(t) ]$, can
also be found from the exact master equation:
\begin{align}
\label{rho1EOM} \frac{d}{dt}\rho^{(1)}_{ij}(t)  = \sum_{m} \{
[\dot{\boldsymbol{u}}(t,t_{0})\boldsymbol{u}^{-1}(t,t_{0})]_{im}
\rho^{(1)}_{mj}(t) \notag \\  + \rho^{(1)}_{im}(t)
[\dot{\boldsymbol{u}}(t,t_{0})\boldsymbol{u}^{-1}(t,t_{0})]^{\dag}_{mj}
\} + \widetilde{\gamma}_{ij}(t) \notag \\   - i f'_{i}(t) \langle
a^{\dag}_{j} (t) \rangle + i \langle a_{i} (t) \rangle
{f'_j}^{\dag}(t) \ .
\end{align}
The solution of the above equation of motion can also be obtained
explicitly in terms of the functions $\boldsymbol{u}(t,t_{0})$,
$\boldsymbol{v}(t,t)$ and the driving-induced field
$\boldsymbol{y}(t)$:
\begin{align}
\rho^{(1)}_{ij}(t)  =  \sum_{mn} u_{im}(t,t_{0})
\rho^{(1)}_{mn}(t_{0})u^*_{ni}(t,t_{0}) + v_{ij}(t,t) \notag
\\  + \langle a_{i}(t) \rangle y^*_{j}(t) + y_{i}(t) \langle
a^{\dag}_{j} (t) \rangle \ - y_{i}(t) y^*_{j}(t) .  \label{rho1sol}
\end{align}
The first term in the above solution comes from the initial photonic
distribution in the resonators. The second term is the effect
induced by the thermal fluctuation in the waveguides (see the
solution of $\boldsymbol{v}(t,t)$ by Eq.~(\ref{vtt})). The remaining
terms are the contribution coming from the driving fields. Besides,
the $i$th diagonal element of the single particle reduced density
matrix gives the average photon number of the mode $i$ in the
resonators, i.e.
\begin{align}
\rho^{(1)}_{ii}(t) ={\rm tr}_s[ a^{\dag}_{i} a_{i} \rho(t) ] =
n_{i}(t), \label{photonm}
\end{align}
which measures the photonic intensity of the $i$th mode in the
resonators. The photonic coherence and the photonic intensity of the
driven resonators, given respectively by Eqs.~(\ref{aEOMsol}) and
(\ref{photonm}), play the important role for the photonic transport
in photonic networks, as we shall discuss in the next section.

\section{Photonic transport in photonic networks}

In this section, we are going to derive the exact photocurrent
flowing from the resonators into each waveguide. Connection between
the photocurrent and the master equation of the reduced density
matrix is given, which explicitly demonstrates the intimacy between
quantum coherence and quantum transport in nonequilibrium dynamics.
Also, the relation between the transport theory obtained here with
the transport theory based on the Keldysh's nonequilibrium Green
function technique is discussed and the powerfulness of the present
theory is given.

\subsection{Photocurrent in each waveguide channel}

To find the photocurrent flowing from the resonators to each
waveguide using the master equation, it will be more convenient to
reexpress the master equation, Eq.~(\ref{ME}), as follows:
\begin{equation}
\label{ME2} \frac{d\rho(t)}{dt} = -i[H_{S}(t), \rho(t)] +
\sum_{\alpha}[\mathcal {L}_{\alpha}^{+}(t) + \mathcal
{L}_{\alpha}^{-}(t)]\rho(t) \ .
\end{equation}
Here $H_{S}(t)$ is the original Hamiltonian of Eq.~(\ref{Hs}) for
the driven resonators. $\mathcal {L}_{\alpha}^{+}(t)$ and $\mathcal
{L}_{\alpha}^{-}(t)$ are superoperators acting on the reduced
density matrix, induced by the coupling to the waveguides. They are
given by:
\begin{subequations}
\label{superop}
\begin{align}
\mathcal {L}_{\alpha}^{+}(t)& \rho(t) = \sum_{ij} \{ \lambda_{\alpha
ij}(t)[ a_{j} \rho(t) a^{\dag}_{i} - \rho(t) a_{j} a^{\dag}_{i} ]
\notag \\ & - \kappa_{\alpha ij}(t) a^{\dag}_{i} a_{j} \rho(t) -
if_{\alpha
i}(t) a_{i}^{\dag} \rho(t) + {\rm H.c.} \} \ , \\
\mathcal {L}_{\alpha}^{-}(t) & \rho(t) = \sum_{ij} \{
\lambda_{\alpha ij}(t)[ a^{\dag}_{i} \rho(t) a_{j} - a^{\dag}_{i}
a_{j} \rho(t)] \notag \\ & + \kappa_{\alpha ij}(t) a_{j} \rho(t)
a^{\dag}_{i} + if_{\alpha i}(t) \rho(t) a_{i}^{\dag} + {\rm H.c.} \}
\end{align}
\end{subequations}
The superoperators $\mathcal {L}_{\alpha}^{+}(t)$ and $\mathcal
{L}_{\alpha}^{-}(t)$ are intimately related to the photocurrent
through the waveguide $\alpha$, as we will see next.

The photocurrent flowing from the resonators into the waveguide
$\alpha$ is defined in the Heisenberg picture as:
\begin{align}
\label{photocurrentdef} I_{\alpha}(t) & \equiv   \frac{d \langle
N_{\alpha}(t) \rangle}{dt} = - i \langle[ N_{\alpha}(t),
H(t)]\rangle \notag \\ & = i\sum_{ki}[ V_{i \alpha k}(t) \langle
a^{\dag}_{i}(t) c_{\alpha k}(t) \rangle - V^{*}_{i \alpha k}(t)
\langle a^{\dag}_{i}(t) c_{\alpha k}(t) \rangle ] \ ,
\end{align}
where $N_\alpha=\sum_k c^\dag_{\alpha k}c_{\alpha k}$ is the photon
number operator of the waveguide $\alpha$. Furthermore, consider the
single particle reduced density matrix in the Heisenberg picture:
$\rho^{(1)}_{ij}(t) = {\rm tr}_s[ a^{\dag}_{j} a_{i} \rho(t)
]=\langle a^{\dag}_{j}(t) a_{i}(t)\rangle$, we find that
\begin{equation}
\label{Heisengbergrho1} \frac{d\boldsymbol{\rho}^{(1)}(t)}{dt} =
-i[\boldsymbol{\omega}, \boldsymbol{\rho}^{(1)}(t)] +
\boldsymbol{\cal S}(t) - \sum_{\alpha}\boldsymbol{\cal
I}_{\alpha}(t) \ ,
\end{equation}
where $\boldsymbol{\cal I}_{\alpha}(t)$ and $\boldsymbol{\cal S}(t)$
are the current matrix of the waveguide $\alpha$ and the source
matrix of the driven resonators:
\begin{subequations}
\begin{align}
{\cal S}_{ij}(t) & = i \langle a_{i}(t) \rangle f^*_{j}(t) - i
f_{i}(t) \langle a^{\dag}_{j}(t) \rangle \ , \label{source}
\\
{\cal I}_{\alpha ij}(t) & = i\sum_{k}[ V_{i \alpha k}(t) \langle
a^{\dag}_{j}(t) c_{\alpha k}(t) \rangle - V^{*}_{j \alpha k}(t)
\langle a^{\dag}_{i}(t) c_{\alpha k} \rangle ]  \label{currentma}.
\end{align}
\end{subequations}
Comparing Eqs.~(\ref{photocurrentdef}) and (\ref{currentma}), one
can see that the trace of the current matrix is just the
photocurrent flowing from the resonators into the waveguide
$\alpha$: $I_\alpha(t)= {\rm Tr}[\boldsymbol{\cal I}_{\alpha}(t)]$.
Note that ${\rm Tr}$ is the trace over the $N \times N$ matrix of
the $N$ single modes in the resonators, while ${\rm tr}_s$ and ${\rm
tr}_E$ used before denote the traces over all the quantum states of
the resonators and the waveguides, respectively.

On the other hand, the equation of motion for the single particle
reduced density matrix can also be obtained directly from the master
equation, Eq.~(\ref{ME2}). The result is
\begin{align}
\label{rho1EOM2} \frac{d}{dt}\rho^{(1)}_{ij}(t) = &
-i[\boldsymbol{\omega}, \boldsymbol{\rho}^{(1)}(t) ]_{ij} +
S_{ij}(t) \notag \\ & + \sum_{\alpha}{\rm tr}_s[a^{\dag}_{j} a_{i}
[\mathcal {L}^{+}_{\alpha}(t) + \mathcal {L}^{-}_{\alpha}(t)]\rho(t)
] \ .
\end{align}
Comparing Eqs.~(\ref{Heisengbergrho1}) and (\ref{rho1EOM2}) for the
single particle reduced density matrix, with the help of
Eqs.~(\ref{superop}) and (\ref{coeff2}), we obtain the explicit
formula for the photocurrent matrix:
\begin{equation}
\label{currentop} \boldsymbol{\cal I}_{\alpha}(t) =
\int_{t_{0}}^{t}d\tau
\big\{\mathbf{g}_{\alpha}(t,\tau)\boldsymbol{\rho}^{(1)}(\tau, t) -
\widetilde{\mathbf{g}}_{\alpha}(t,\tau)\bar{\boldsymbol{u}}(\tau, t)
+ {\rm H.c.} \big\}
\end{equation}
and $\boldsymbol{\rho}^{(1)}(\tau, t)$ is the generalized
correlation function which is given explicitly by:
\begin{align}
\label{rho1taut}  \rho^{(1)}_{ij}(\tau, t) & =  \sum_{mn}
u_{im}(\tau,t_{0}) \rho^{(1)}_{mn}(t_{0})u^*_{ni}(t,t_{0}) +
v_{ij}(\tau,t) \notag \notag \\ &+ y_{i}(\tau)y^*_{j}(t) + \sum_{m}[
u_{im}(\tau,t_{0})\langle a_{m}(t_{0}) \rangle y^*_{j}(t) \notag \\
&~~~~~~~~+ y_{i}(\tau)\langle a^{\dag}_{m}(t_{0}) \rangle
u^*_{mj}(t,t_{0})] \ .
\end{align}
 Then the photocurrent flowing into
the waveguide $\alpha$ is simply given by:
\begin{equation}
\label{photocurrent} I_{\alpha}(t) = 2 {\rm Re}{
\int_{t_{0}}^{t}d\tau {\rm
Tr}[\mathbf{g}_{\alpha}(t,\tau)\boldsymbol{\rho}^{(1)}(\tau, t)
-\widetilde{\mathbf{g}}_{\alpha}(t,\tau)\bar{\boldsymbol{u}}(\tau,
t)] } .
\end{equation}
It shows that the photocurrent  is completely determined by the
time-correlation functions $\mathbf{g}_{\alpha}(t,\tau)$ and
$\widetilde{\mathbf{g}}_{\alpha}(t,\tau)$ of the waveguides and the
propagating and correlation functions $\bar{\boldsymbol{u}}(\tau,
t)$ and $\boldsymbol{\rho}^{(1)}(\tau, t)$ of the driven resonators.
The time-correlation functions characterize the non-Markovian memory
structure of photon transfer between resonators and waveguides,
while the propagating and correlation functions depict completely
the photon coherence and photon correlation of the resonators under
the control of external driving fields.

In fact, Eq.~(\ref{Heisengbergrho1}) is a generalized quantum
continuous equation. By tracing over the equation of motion for the
single particle reduced density matrix, we have
\begin{equation}
\frac{d N}{dt} = S(t) - \sum_{\alpha} I_{\alpha}(t) \ . \label{qce}
\end{equation}
Here $N(t) = {\rm Tr}[\boldsymbol{\rho}^{(1)}(t)]$ is the total
photon number in the resonators, $S(t) = {\rm Tr}[\boldsymbol{\cal
S}(t) ]$ is the source coming from the driving fields, and
$I_{\alpha}(t) = {\rm Tr}[ \boldsymbol{\cal I}_\alpha(t) ]$ is the
photocurrent flowing into the waveguides $\alpha$. Eq.~(\ref{qce})
tells that the increase of the photon number in the resonators
equals to the receiving photons from the driving field subtracting
the lessening photons flowing into waveguides.

\subsection{Relations to the Keldysh's nonequilibrium Green function technique}

As one seen from Eq.~(\ref{photocurrent}), the photocurrent is
completely determined by the time-correlation functions
$\mathbf{g}_{\alpha}(t,\tau)$ and
$\widetilde{\mathbf{g}}_{\alpha}(t,\tau)$ of the waveguides  plus
the propagating and correlation functions
$\bar{\boldsymbol{u}}(\tau, t)$ and $\boldsymbol{\rho}^{(1)}(\tau,
t)$ of the driven resonators. In fact, we have shown
\cite{Jin10083013} that the functions $\boldsymbol{u}(\tau, t_{0})$,
$\bar{\boldsymbol{u}}(\tau,t)$, $\boldsymbol{\rho}^{(1)}(\tau, t)$
are related to the retarded, advanced and lesser Green functions of
the resonators in the Keldysh's nonequilibrium formulism
\cite{Sch61407,Cho851,Hau98}:
\begin{subequations}
\begin{align}
& u_{ij}(t_1, t_2)=\theta(t_1-t_2)\langle[a_i(t_1),
a^\dag_j(t_2) ] \rangle \equiv iG^r_{ij}(t_1, t_2) , \\
& \bar{u}_{ij}(t_1, t_2)= \theta(t_2-t_1)\langle[a_i(t_1),
a^\dag_j(t_2)] \rangle \equiv-iG^a_{ij}(t_1, t_2) , \\
& \rho^{(1)}_{ij}(t_1, t_2)= \langle a^\dag_j(t_2) a_i(t_1)\rangle
\equiv -iG^<_{ij}(t_1, t_2)  .
\end{align}
\end{subequations}
While, the  time-correlation functions $\mathbf{g}_{\alpha}(t,\tau)$
and $\widetilde{\mathbf{g}}_{\alpha}(t,\tau)$ correspond to the
retarded and lesser self-energy functions arose from the couplings
between the resonators and the waveguides:
\begin{subequations}
\begin{align}
&\mathbf{g}_{\alpha ij}(t_1, t_2)= i\Sigma^r_{\alpha ij}(t_1, t_2) , \\
&\widetilde{\mathbf{g}}_{\alpha ij}(t_1, t_2)=-i\Sigma^<_{\alpha
ij}(t_1, t_2) .
\end{align}
\end{subequations}
The explicit form of these self-energy functions is given by
Eq.~(\ref{t-correlations}) or (\ref{spectralf}) which is rather
simple. In the Keldysh's nonequilibrium Green function technique,
quantum transport theory is completely determined by the retarded,
advanced and lesser Green functions.

Explicitly, Eq.~(\ref{sol2b}) for $\boldsymbol{u}(\tau, t_{0})$
obtained in the last section can be rewritten as:
\begin{align}
\Big\{i\frac{d}{d\tau} - \boldsymbol{\omega}\Big\} &
\boldsymbol{G}^r(\tau, t_{0})= \delta(\tau-t_0) \notag \\
& + \int_{t_{0}}^{\tau} \mathbf{\Sigma}^r(\tau,
\tau')\boldsymbol{G}^r(\tau', t_{0})d\tau'
\end{align}
which is just the standard Dyson equation for the retarded Green
function. The advanced Green function obeys the relation:
$\boldsymbol{G}^a(t_1, t_2)=[\boldsymbol{G}^r(t_2, t_1)]^\dag$, see
Eq.~(\ref{ubars}). The central and also the most difficult part in
the Keldysh's nonequilibrium Green function technique is the
calculation of the lesser Green function $\boldsymbol{G}^<(\tau,t)$.
The lesser Green function $\boldsymbol{G}^<(t_1, t_2)$ fully
determines the quantum kinetic theory of nonequilibrium system. From
Eq.~(\ref{rho1taut}), we have obtained already the exact analytical
solution of the lesser Green function:
\begin{align} \label{lessGtaut}
\boldsymbol{G}^< (\tau,& t)  =  \boldsymbol{G}^r(\tau,t_{0})
\boldsymbol{G}^<(t_{0},t_0)\boldsymbol{G}^a(t_0,t)
+ i\boldsymbol{y}(\tau)\boldsymbol{y}^\dag(t)  \notag \\
& - \boldsymbol{G}^r(\tau,t_{0})\langle \boldsymbol{a}^\dag(t_{0})
\rangle \boldsymbol{y}^\dag(t) + \boldsymbol{y}(\tau) \langle
\boldsymbol{a}(t_{0})
\rangle\boldsymbol{G}^a(t_0,t)\notag \\
& + \int^\tau_{t_0} d\tau_1 \int_{t_0}^{t }d\tau_2 ~
\boldsymbol{G}^r(\tau, \tau_1)\boldsymbol{\Sigma}^<(\tau_1,\tau_2)
\boldsymbol{G}^a(\tau_2,t),
\end{align}
where $\boldsymbol{G}^<(t_{0},t_0)=i\boldsymbol{\rho}^{(1)}(t_0)$ is
the initial photon distribution in the resonators, $\langle
\boldsymbol{a}(t_{0}) \rangle $ is the initial resonator fields, and
$\boldsymbol{y}(t)$ is the driving-induced resonator field given by
Eq.~(\ref{difield}).

Comparing with the electron transport in mesoscopic systems, one
always has $\langle \boldsymbol{a}^\dag(t_{0}) \rangle =\langle
\boldsymbol{a}(t_{0}) \rangle =0$ for electrons. Also, one usually
ignores the first term in the standard Green function calculation by
taking $t_0 \rightarrow -\infty$ which loses the information of the
initial state dependence in quantum transport, an important effect
on non-Markovian memory dynamics. Besides, the external driving
fields applied to the leads and gates in mesoscopic systems are
embedded into the spectral densities or the energy levels in the
central region so that no extra driving-induced field is produced,
namely $\boldsymbol{y}(t)=0$. Thus, the resulting lesser Green
function obtained in the mesoscopic electron transport contains only
the last term in Eq.~(\ref{lessGtaut}) \cite{Hau98}. In other words,
Eq.~(\ref{lessGtaut}) gives the exact and general solution for the
lesser Green function in photonic systems. With the above relations
and solutions, the photocurrent, Eq.~(\ref{photocurrent}), can be
re-expressed as:
\begin{equation}
\label{photocurrent1} I_{\alpha}(t) = 2 {\rm Re}{
\int_{t_{0}}^{t}d\tau {\rm
Tr}[\mathbf{\Sigma}^r_{\alpha}(t,\tau)\boldsymbol{G}^<(\tau, t)
+\mathbf{\Sigma}^<_{\alpha}(t,\tau)\boldsymbol{G}^a(\tau, t)] } .
\end{equation}
This reproduces the standard transport current in the Keldysh's
nonequilibrium Green function technique that has been widely used in
the investigation of various electron transport phenomena in
mesoscopic systems, although most of the previous works used only
the special solution of the lesser Green function, as we just
discussed above.

In conclusion, a full quantum transport theory for photonic dynamics
in photonic network has been established based on the Feynman-Vernon
influence functional approach. In the literature, the investigation
of quantum transport used mainly the Keldysh's nonequilibrium Green
function technique. The Keldysh's nonequilibrium Green function
technique has the advantage of treating the lesser Green function in
a complicated system by the assumption of adiabatically switching on
the many-body correlations. This allows one to trace back the
initial time $t_0 \rightarrow -\infty$, which provides a great
simplification for practical evaluation but meantime it lacks a
proper treatment of transient phenomena. The Feynman-Vernon
influence functional aims to address dissipative dynamics of an open
system in terms of the reduced density matrix (an arbitrary quantum
state). The master equation derived from the influence functional
explicitly determines, by definition, the temporal evolution of an
initially prepared state. For the photonic networks considered in
this paper, the many-photon correlations are less important and can
be ignored so that we are able to obtain the exact master equation
where all the non-Markovian memory effects are encoded into the
time-dependent coefficients in the master equation. It turns out
that these time-dependent coefficients are determined indeed by the
retarded and less Green functions in Keldysh's formulism, as we have
just shown. Therefore, all the advantages of the nonequilibrium
Green function technique are maintained in our theory but the
difficulty in addressing the transient dynamics in Green function
technique is avoided in terms of the master equation. As a result,
we unify two fundamental nonequilibrium approaches, the
Schwinger-Keldysh nonequilibrium Green function technique and the
Feynman-Vernon influence functional approach, together to make our
theory more powerful in the study of the transient transport
phenomena in nonequilibrium photonic systems.

\section{Analytical and Numerical illustration}

\subsection{Waveguide as a tight-binding model}
In this section, we apply the theory developed in the previous
sections to a photonic circuit in photonic crystals as an
illustration.  One of the physical realization for the photonic
network described in Sec.II is the system of one or a few
nanocavities coupled to many waveguides in the photonic crystals, as
schematically shown by Fig.~\ref{SysDem}. Here a nanocavity can be
considered as a point defect created in photonic crystals as a
resonator \cite{HQcavity}. Its frequency can easily be tuned to any
value within the band gap by changing the size or the shape of the
defect. While, a waveguide in photonic crystals can also be
considered as a series of coupled point defects in which photon
propagates due to the coupling of the adjacent defects
\cite{SlowLightObs,CROWproposal}. By changing the modes of the
resonators and the coupling configuration through various techniques
\cite{thermotunePBG,electroopticcontrol,intoptofluodic,mectune}, the
transmission properties of the waveguide can also be manipulated. In
principle, one can solve Eq.~(\ref{normm}) to obtain the dispersion
relations of the waveguides and the couplings between the cavities
and waveguides \cite{PC,spg}. Here we may treat the waveguide as a
tight-binding model, namely the Hamiltonian of the waveguide and the
coupling Hamiltonian between the nanocavities and the waveguide can
be expressed explicitly as \cite{Wu1018407}:
\begin{subequations}
\label{wgHam}
\begin{align}
& H_{E\alpha} = \sum_n \omega_{\alpha} c_{\alpha n}^{\dag} c_{\alpha
n} - \sum_{n=1} \xi_{\alpha} (c_{\alpha n}^{\dag} c_{\alpha n+1} +
{\rm H.c.}) \ , \\
& H_{T\alpha}= \xi_{i \alpha n_i} (a^{\dag }_i c_{\alpha n_i} + {\rm
H.c.}) \ . \label{hh}
\end{align}
\end{subequations}
where $\xi_{\alpha}$ is the hopping rate between adjacent resonator
modes within the waveguide $\alpha$, and it is experimentally
turnable. $\xi_{i \alpha n_i}$ is the coupling constant of the $i$th
nanocavity coupled to the $n$th resonator in the waveguide $\alpha$,
which is also controllable by changing the geometrical parameters of
the defect cavity and the distance between the cavities and the
waveguide \cite{cavity with CROW-1}.

Furthermore, making the Fourier transform for semi-infinite
waveguide,
\begin{align}
c_{\alpha k}=\sqrt{\frac{2}{\pi}} \sum_{n=1}^{\infty}
\sin(nk)c_{\alpha n} \ ,
\end{align}
we can transform the Hamiltonian of Eq.~(\ref{wgHam}) from the
spatial space into the wavevector space with the result:
\begin{subequations}
\label{wgHam1}
\begin{align}
& H_{E\alpha} = \sum_k \omega_{\alpha k} c_{\alpha k}^{\dag}
c_{\alpha k} \ , \\
& H_{T\alpha}= \sum_k \big(V_{i\alpha k} a^{\dag}_i c_{\alpha k} +
{\rm H.c.}\big) \ ,
\end{align}
\end{subequations}
where $0 \leq k \leq \pi$, and $\omega_{\alpha k}$ and $V_{i \alpha
k}$ are given by:
\begin{eqnarray}
\omega_{\alpha k} =  \omega_{\alpha}-2\xi_{\alpha}\cos(k), \ V_{i
\alpha k}=\sqrt{\frac{2}{\pi}}~\xi_{i \alpha n_i} \sin(n_i k),
\label{vk}
\end{eqnarray}
and $c_{\alpha k}^{\dag},c_{\alpha k}$ are the creation and
annihilation operators of the corresponding Bloch modes in the
waveguide $\alpha$.

More specifically, Fig.~\ref{SysDem}(a) consists of $N$ nanocavities
coupled to one waveguide at different sites $\{n_i\}$. Then the
total Hamiltonian of the system can be rewritten explicitly as:
\begin{align}
H_a(t) = &  \sum_{i=1}^N \big(\omega_i a^\dag_i a_i + f_i(t)
a^\dag_i
+ f^*_i(t)a_i\big) \notag \\
& + \sum_{k} \omega_{k} c_{k}^{\dag} c_{k} + \sum_{i k} \big(V_{i k}
a^{\dag}_i c_{k} + {\rm H.c.}\big) \ ,
\end{align}
where $\omega_{k} =  \omega_{0}-2\xi_0 \cos(k)$, $V_{i
k}=\sqrt{\frac{2}{\pi}}~\xi_{i n_i} \sin(n_i k)$. The network
presented in Fig.~\ref{SysDem}(b) shows a nanocavity (at the center)
coupled to $M$ individual waveguides. The corresponding Hamiltonian
is
\begin{align}
H_b(t) & =   \omega_c a^\dag a + f(t) a^\dag
+ f^*(t)a  \notag \\
& + \sum_{\alpha k} \omega_{\alpha k} c_{\alpha k}^{\dag} c_{\alpha
k} + \sum_{\alpha k} \big(V_{\alpha k} a^{\dag} c_{\alpha k} + {\rm
H.c.}\big) \ ,
\end{align}
where the band structure of the waveguides and the coupling
constants are given by Eq.~(\ref{vk}) with $n_i=1$. These physical
realizations specify the model Hamiltonian of Eq.~(\ref{HM}).

\begin{figure}
\centering
\includegraphics[width = 6 cm]{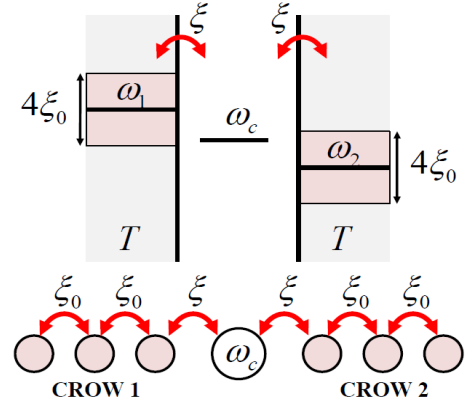}
\caption{A nanocavity with frequency $\omega_c$ couple to two
coupled resonator optical waveguides (CROWs).} \label{Sys}
\end{figure}

As an illustration, we consider here simply a driven nanocavity
coupled to two coupled resonator optical waveguides (CROWs) in
photonic crystals, as plotted in Fig.~\ref{Sys}, and then calculate
analytically and numerically the photonic transport phenomena. The
Hamiltonian of this simple photonic circuit is :
\begin{align}
H & = \omega_{c} a^{\dag}a  + (E_{0} e^{-i\omega_{d} t} a^{\dag} +
E_{0} e^{i\omega_{d} t} a ) \notag \\ & + \sum_{\alpha = 1}^2
\sum_{k} \omega_{\alpha k} c_{\alpha k}^{\dag} c_{\alpha k} +
\sum_{\alpha = 1}^{2} \sum_{k} ( V_{\alpha k} a^{\dag} c_{\alpha k}
+ V^{*}_{\alpha k} a c^{\dag}_{\alpha k} ) \ ,
\end{align}
where $E_{0}$ is the strength of the external driving fields in
frequency $\omega_{d}$. The frequency $\omega_{\alpha k}$ and
coupling constant $V_{\alpha k}$ are given by Eq.~(\ref{vk}) with
$n_i=1$. The corresponding spectral densities are given by
$J_\alpha(\omega)=2\pi g_\alpha(\omega)|V_\alpha(\omega)|^2$ where
$g(\omega)$ is the density of state of the waveguide $\alpha$ and
$V_\alpha(\omega)$ is the coupling between the cavity and the
waveguide $\alpha$. They can be calculated directly from
Eq.~(\ref{vk}):
\begin{subequations}
\begin{align}
& g_\alpha(\omega) = \frac{dk}{d\omega}=\frac{1}{
\sqrt{4\xi_\alpha^2-(\omega-\omega_\alpha)^2}}, \\
& V_\alpha(\omega) = \frac{1}{\sqrt{2\pi}}\Big(
\frac{\xi_{c\alpha}}{\xi_\alpha}\Big)\sqrt{4\xi_\alpha^2-(\omega-\omega_\alpha)^2},
\label{vks}
\end{align}
\end{subequations}
with $\omega_\alpha - 2 \xi_\alpha < \omega < \omega_\alpha + 2
\xi_\alpha$. Then the spectral density can be explicitly written as
\begin{align}
J_{\alpha}(\omega) = \left\{
\begin{array}{ ll }
\eta_{\alpha}^{2}\sqrt{4\xi_{\alpha}^{2} - (\omega -
\omega_{\alpha})^{2}}
 \ , & |\omega - \omega_{\alpha}| \leq 2 \xi_{\alpha} \  \\
0 \ , & |\omega - \omega_{\alpha}| > 2 \xi_{\alpha} \\
\end{array} \right.  \label{spde}
\end{align}
with $\eta_{\alpha} = \xi_{c\alpha} / \xi_{\alpha}$. In practical,
$\xi_\alpha \ll \omega_\alpha$, namely the waveguide has a very
narrow band.

%===================================================================
\subsection{Analytical solutions in the weak coupling regime}
First, we shall discuss the cavity photonic dynamics. Consider the
case where the nanocavity is initially empty: $\langle a(t_0)
\rangle=0$ and $n(t_0)=0$. The cavity field and the cavity photon
number
 at a later time, i.e. Eqs.~(\ref{aEOMsol}) and
(\ref{rho1sol}), become
\begin{subequations}
\label{cavity}
\begin{align}
\label{solaemp}   & \langle a(t) \rangle = y(t) \ , \\
\label{solrho1emp} & n(t) = \langle a^{\dag}(t) a(t) \rangle  =
v(t,t) + |y(t)|^{2} \ .
\end{align}
\end{subequations}
Here $y(t)$ is the driving-induced field given by
Eq.~(\ref{difield}) and $v(t,t)$ is the photon correlation due to
the thermal fluctuation in the waveguides, given by Eq.~(\ref{vtt}).
Only at zero temperature, we have $v(t,t)=0$. Then the cavity photon
number equals to the absolute square of the cavity field, i.e. $n(t)
= |\langle a(t) \rangle|^{2}$.

To see the dynamics of cavity field controlled through the driving
field, we shall consider first the cavity decoupled to waveguides,
i.e. $\xi_{1c} = \xi_{2c} = 0$. In this situation, the photonic
propagating function in the cavity is simply given by $u(t, t_{0}) =
e^{-i\omega_{c}(t-t_{0})}$ and the photon correlation function
$v(t,t) = 0$. Then, the cavity field and cavity photon number equal
to $y(t)$ and $|y(t)|^{2}$, respectively, with
\begin{align}
\label{insulatedy}  & y(t) =
\begin{cases}
\frac{E_{0}}{\omega_{d} - \omega_{c}}[e^{-i\omega_{d}(t-t_{0})} -
e^{-i\omega_{c}(t-t_{0})}] & \omega_{d} \neq \omega_{c} \\ & \\
E_{0}(t-t_{0})e^{-i(\omega_{c}(t-t_{0}) + \frac{\pi}{2})} &
\omega_{d} = \omega_{c}
\end{cases} \ .
%& |y(t)|^{2} =
%\begin{cases}
%\frac{E_{0}^{2}}{(\omega_{D} -
%\omega_{c})^{2}}\sin^{2}((\frac{\omega_{d}
%-\omega_{c}}{2})(t-t_{0})) & \omega_{d} \neq \omega_{c} \\
%E_{0}^{2}(t-t_{0})^{2} & \omega_{d} = \omega_{c}
%\end{cases}
\end{align}
It shows that the cavity field is a coherent superposition of the
cavity mode with the external driving field. If the driving field
frequency differs from the cavity mode, the interference between the
cavity mode and the driving field causing the photons jump in and
out of the cavity with the Rabi frequency $
(\omega_{d}-\omega_{c})/2$. When the driving field is in resonance
with the cavity mode, the driving field would be completely absorbed
into the cavity, and the cavity field amplitude increases linearly
in time without an interference oscillation.

When the cavity couples to waveguides, generally it is not easy to
find the analytical solution for $u(t,t_0)$ and $v(t,t)$. One has to
solve Eqs.~(\ref{sol2a}) and (\ref{vtt})  numerically to understand
the photonic dynamics of the driven cavity. However, in the weak
coupling regime, the memory effect is negligible so that the
Born-Markov approximation can be applied \cite{Xio10012105}. Then
the photonic propagating function and the correlation function in
weak coupling limit can be approximated (see Appendix B) by
\begin{subequations}
\begin{align}
& \label{BMsolu0} u(t,t_{0}) \simeq e^{-(i\omega'_{c} +
\kappa)(t-t_{0})} \ , \\
& \label{BMsolv0} v(t,t) \simeq \bar{n}(\omega'_c,T)[1 -
e^{-2\kappa(t-t_{0})}] \ ,
\end{align}
\end{subequations}
where $\omega'_c=\omega_c+ \sum_\alpha\delta \omega_\alpha$ is a
renormalized cavity frequency with frequency shift $\delta
\omega_\alpha= {\cal P}[\int \frac{d\omega}{2\pi}
\frac{J_\alpha(\omega)}{\omega_c-\omega}]$. The damping rate $\kappa
=\sum_\alpha\kappa_\alpha$ with $\kappa_\alpha=J_\alpha(\omega_c)/2$
and the average photon number of the two waveguides
$\bar{n}(\omega'_c,T) =\sum_\alpha J_\alpha(\omega'_c)
n_\alpha(\omega'_c)/2\kappa$.

On the other hand, to solve the driving-induced field $y(t)$ in the
BM limit, special care needs to be taken since the driving field has
its own character frequency $\omega_d$ which is usually different
from the cavity mode frequency $\omega_c$. Therefore, instead of
applying directly the BM solution of Eq.~(\ref{BMsolu0}) into
Eq.~(\ref{difield}), we have to use different character frequencies
$\omega_c$ and $\omega_d$ for the homogeneous and inhomogeneous
solutions of Eq.~(\ref{sol2d}) to find the BM limit of the
driving-induced field $y(t)$. The detailed derivation is also given
in Appendix B. The result is nontrivial:
\begin{align}
y(t) \simeq \frac{E_{0}\exp{(-i\phi)}}{\sqrt{(\omega_{d} -
\tilde{\omega}_{c})^2 + \tilde{\kappa}^2}} [& e^{-i\omega_{d}(t-t_{0})} \notag \\
&- e^{-(i\omega'_{c} + \kappa)(t-t_{0})} ] ,
\end{align}
where
$\phi=\tan^{-1}\frac{\tilde{\kappa}}{\omega_d-\tilde{\omega}_c}$,
$\tilde{\omega}_c = \omega_{c} + \sum_{\alpha=1}^{2}\delta
\tilde{\omega}_\alpha$ with the driving field induced frequency
shift $\delta \tilde{\omega}_\alpha= {\cal P}\int_{0}^{\infty}
\frac{d\omega}{2\pi} \frac{J_{\alpha}(\omega)}{ \omega_d -\omega}$.
$\tilde{\kappa} = \sum_{\alpha=1}^{2}\tilde{\kappa}_\alpha$ with
$\tilde{\kappa}_\alpha=J_{\alpha}(\omega_{d})/2$, located at the
driving frequency rather than the cavity frequency. Correspondingly,
the cavity field and the cavity photon number of Eq.~(\ref{cavity})
in the BM limit become
\begin{subequations}
\label{BMcavity}
\begin{align}
\langle a(t)\rangle & \simeq
\frac{E_{0}\exp{(-i\phi)}}{\sqrt{(\omega_{d} - \tilde{\omega}_{c})^2
+
\tilde{\kappa}^2}} [ e^{-i\omega_{d}(t-t_{0})} \notag \\
&~~~~~~~~~~~~~~~~~~~~~~~~~- e^{-(i\omega'_{c} + \kappa)(t-t_{0})} ] ,\label{BMsola}  \\
n(t)\simeq & ~\bar{n}(\omega'_c,T)[1 - e^{-2\kappa(t-t_{0})}] \notag \\
& + \frac{E_{0}^{2}}{(\omega_{d} - \tilde{\omega}_{c})^{2} +
\tilde{\kappa}^{2}}\big[1 + e^{-2\kappa(t-t_{0})} \notag
\\ & ~~~~~ -2e^{-\kappa(t-t_{0})}
\cos{[(\omega_{d}-\omega'_{c})(t-t_{0})]}\big] . \label{BMsoln}
\end{align}
\end{subequations}
As one see, the cavity field is a coherent superposition of the
external driving field and the damped cavity mode, where the damping
comes from the coupling of the cavity with waveguides that induces
photon dissipation from the cavity into waveguides. This damping
(dissipation) effect leads cavity mode to vanish in the steady
limit. Only the driving field is remained in the cavity with a
modified (amplified) field amplitude
$E_0'=\frac{E_{0}}{\sqrt{(\omega_{d} -
\tilde{\omega}_{c})^2+\tilde{\kappa}^2}}$ and a phase shift $\phi$,
as a feed-back effect from the coupling of the cavity with
waveguides. The photon number in the cavity is then a combination of
the driving-induced field plus a background noise from the thermal
fluctuation of waveguides [$\sim \bar{n}(\omega'_c,T)$]. These
analytical results help us to understand the driven cavity dynamics
in the strong coupling regime. The corresponding numerical solution
will be presented later.

Furthermore, the transport phenomena of photons flowing into the
waveguides can be specified by the photocurrent which can be
simplified as well in the weak coupling limit:
\begin{align}
\label{BMsolI}  I_{\alpha}(t) = &  -2\kappa_{\alpha}
\bar{n}(\omega'_{c})
e^{-2\kappa(t-t_{0})} \notag \\
& + \frac{2E_{0}^{2}\kappa_{\alpha}}{(\tilde{\omega}_{c} -
\omega_{d})^{2} + \tilde{\kappa}^{2}}\Big\{
\frac{\tilde{\kappa}_{\alpha}}{\kappa_\alpha}+ e^{-2\kappa(t-t_{0})} \notag \\
&  - (1 + \frac{\tilde{\kappa}_{\alpha}}{\kappa_\alpha})
e^{-\kappa(t-t_{0})}\cos{[(\omega_{d} - \omega'_{c})(t-t_{0})]}
\notag \\ &  + \frac{\delta \omega_{\alpha} -
\delta\tilde{\omega}_{\alpha}}{\kappa_\alpha}
e^{-\kappa(t-t_{0})}\sin{[(\omega_{d} -
\omega'_{c})(t-t_{0})]}\Big\}.
\end{align}
The first term in Eq.~(\ref{BMsolI}) is the contribution of the
waveguide thermal fluctuation, which would be vanish when the cavity
is equilibrated with the waveguides. The remainders come from the
response of the waveguides to the external driving field through the
cavity, in which the first decay term is due to the dissipation of
the cavity mode, the second and third decay terms are the
thermal-fluctuation-induced decoherence (noise) of the cavity field.
These three contributions together with the
thermal-fluctuation-induced current (the first term) will vanish in
the steady limit. The time-independent term in Eq.~(\ref{BMsolI}) is
the steady photocurrent, which is determined by the cavity mode, the
driving field frequency and the spectrum of the waveguide. When the
driving field frequency lies outside the band of the waveguide
$\alpha$, the steady photoncurrent vanishes because
$\tilde{\kappa}_{\alpha}=J_\alpha(\omega_d)/2= 0$. On the other
hand, the photocurrent flowing into the waveguide $\alpha$ becomes
maximal when the driving field frequency lies at the band center of
the waveguide $\alpha$, i.e. $\omega_{d} = \omega_{\alpha}$, see
Eq.~(\ref{spde}). This result shows that the nontrivial
Born-Markovian solution in the presence of external driving field
gives already a clear picture on the controllability of photonic
transport in the photonic circuit through the external driving
frequency.

%===================================================================
\subsection{Exact numerical solutions in both the weak and strong coupling regimes}
\begin{figure}
\includegraphics[width = 8 cm]{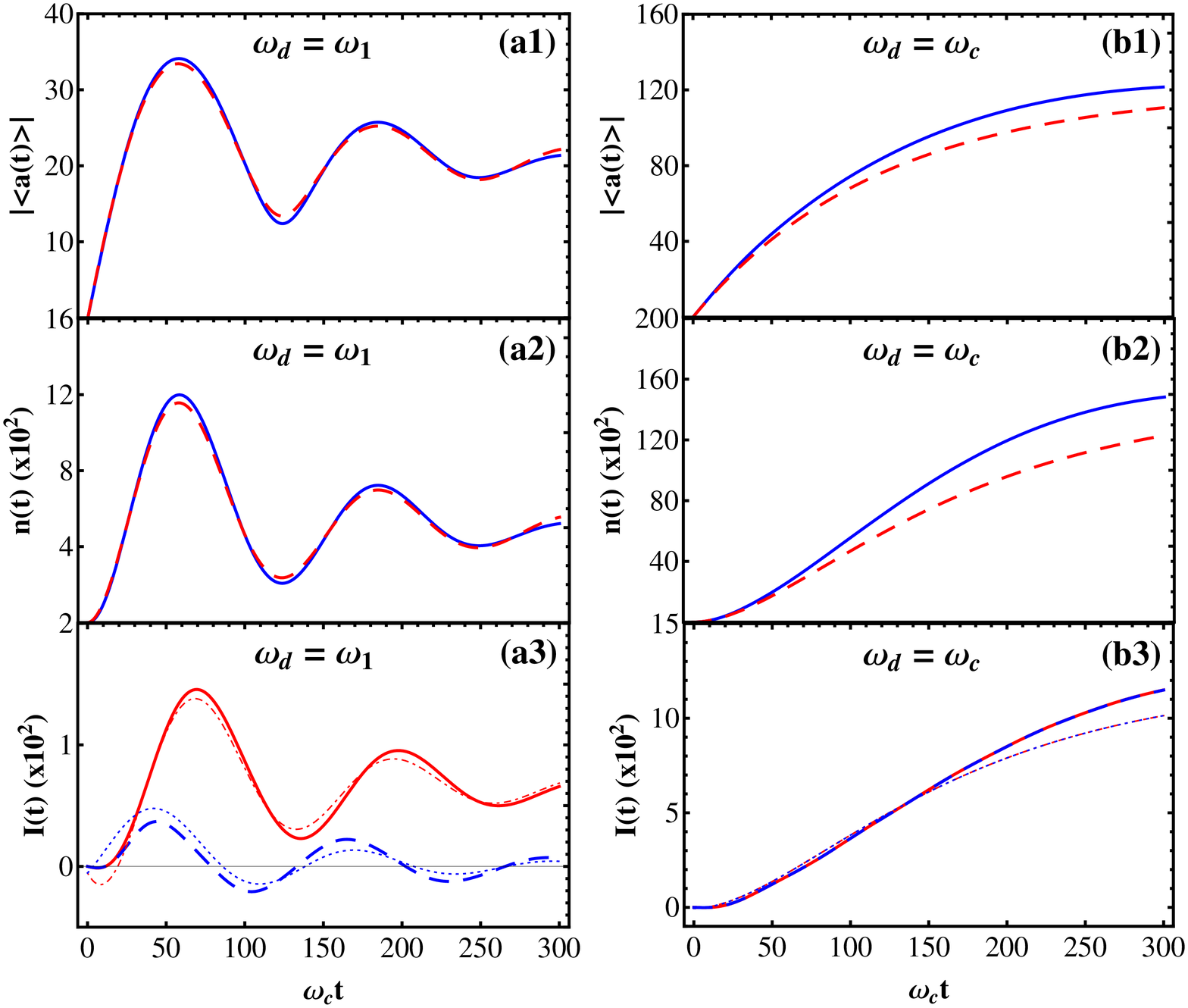}
\caption{Comparison of the analytical solution in the weak coupling
limit with the exact numerical solution at the coupling rate $\eta =
\xi/\xi_0=0.5$ with different driving frequency. (a1)-(a1) and
(a2)-(b2) are the cavity field amplitude and cavity photon number,
respectively, with the BM solution (red dashed line) and the exact
solution (blue solid line). (a3)-(b3) are the photocurrent
$I_{1}(t)$ and $I_{2}(t)$ with the exact solution (red solid line
and blue dashed line) and the BM solution (blue dotted line and red
dotted-dashed line).} \label{cmp}
\end{figure}

Now, we turn to the exact numerical calculation for arbitrary
coupling between the cavity and waveguides. We will focus on the
situation where the frequency of the cavity lies between the band
centers of the two waveguides, i.e. $\omega_{c} = (\omega_{1} +
\omega_{2})/2$. In the numerical calculation, we take
$\omega_{1}=9.5$GHz, $\omega_{2}=10.5$GHz,
$\xi_{1}=\xi_{2}=\xi_{}=0.3$GHz, $\xi_{1 c} = \xi_{2 c} = \xi_c$,
and $E_{0} = 10$GHz. We should first compare the above analytical BM
solution in the weak coupling limit with the exact numerical
solution of the cavity field amplitude, the cavity photon number and
the photoncurrent, given by Eqs.~(\ref{BMcavity}) and (\ref{BMsolI})
and Eqs.~(\ref{cavity}) and (\ref{photocurrent}), respectively. The
result is plotted in Fig.~\ref{cmp} where the coupling rate $\eta =
\xi/\xi_0=0.5$ which belongs to a weak coupling and the BM
approximation is applicable as we have shown in \cite{Wu1018407}.
From Fig.~\ref{cmp}, we see that in the weak coupling limit, the BM
solution is in good agreement with the exact solution, in particular
when the driving frequency $\omega_d=\omega_1$. When the driving
frequency equals to the cavity mode frequency, the deviation between
the exact solution and the BM solution becomes relatively large in
the steady limit. This can be seen from Eq.~(\ref{BMcavity}) where
the amplified cavity field amplitude in the BM limit,
$E_0'=\frac{E_{0}}{\sqrt{\delta
\tilde{\omega}_{c}^2+\tilde{\kappa}^2}}$, becomes sensitive to the
approximation. But the qualitative behavior of the BM solution is
still in good agreement with the exact solution.

\begin{figure}
\centering
\includegraphics[width = 8.5 cm]{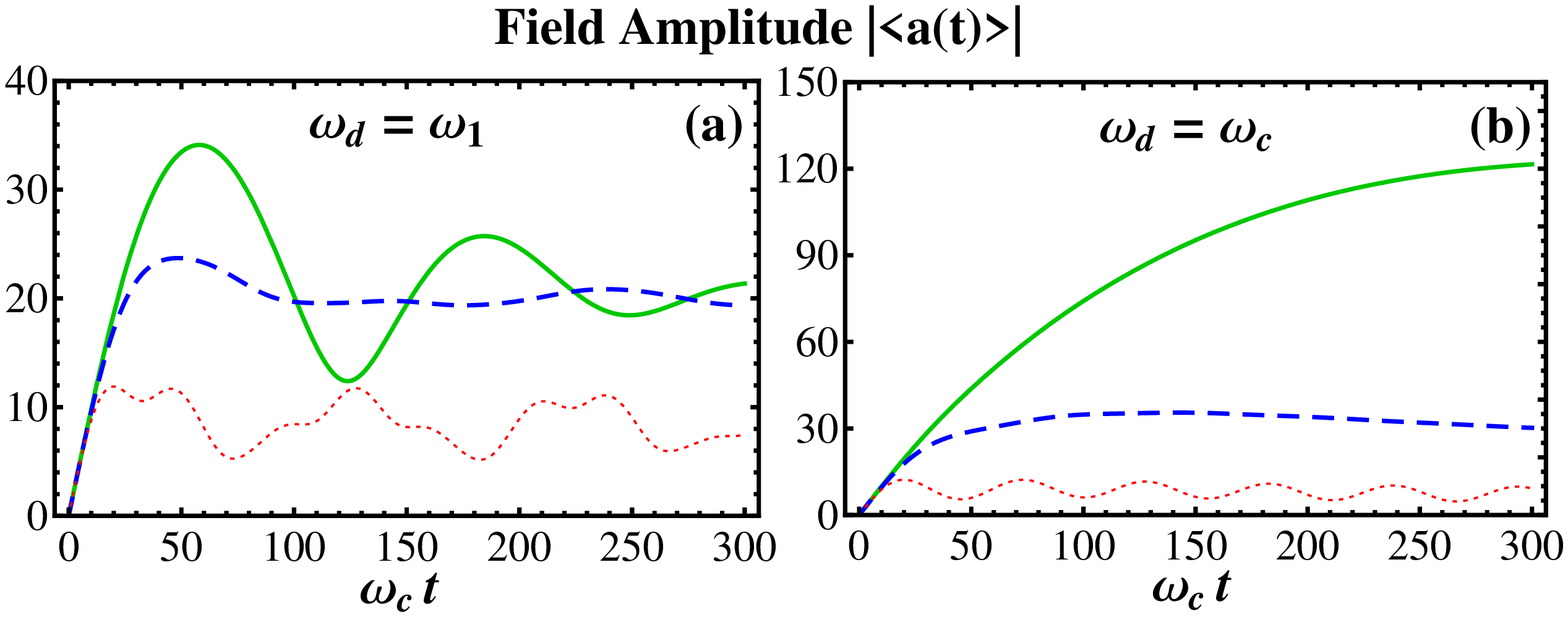}
\caption{The time evolution of the cavity field amplitude with
different coupling strengthes: $\eta=0.5$ (solid green line), $1.0$
(dashed blue line) and $2.0$ (dotted red line). (a) The driving
field frequency equals to the band center of the CROW 1. (b) The
driving field frequency is in resonance to the cavity mode.}
\label{atA}
\end{figure}

Next, we shall present the exact numerical solutions for different
values of the coupling constants between the cavity and the
waveguides, to examine the different photonic transport dynamics in
the weak coupling as well as in the strong coupling regime.
Fig.~\ref{atA} shows the cavity field amplitude of
Eq.~(\ref{cavity}) with different coupling strengths and different
driving field frequencies. Fig.~\ref{atA}(a) and Fig.~\ref{atA}(b)
correspond to the cases of the driving frequency being in resonance
with the band center of CROW 1 and the cavity mode, respectively.
Different curves in each plot correspond to different coupling
strengths between the cavity and waveguides. In the weak coupling
regime, the behavior of the cavity field amplitude highly depends on
the driving field frequency. When the driving field is not in
resonance to the cavity mode, the field amplitude oscillates (as a
result of the superposition of the driving field with the cavity
mode) and decay (due to the dissipation induced by the coupling of
the cavity with waveguides). When the driving field is in resonance
to the cavity mode ($\omega_d=\omega_c$), the field amplitude
increases gradually without oscillation. These numerical results
agree with the BM solution of Eq.~(\ref{BMsola}), as shown in
Fig.~\ref{cmp}. In the strong coupling regime, the field amplitude
keep oscillating without decay in both cases. The absence of the
damping (dissipation) is totally due to the non-Markovian memory
effect, as we have pointed out recently
\cite{Xio10012105,Wu1018407}. The complicated oscillating behavior
(dotted red curve) in Fig.~\ref{atA}(a) is an interference effect
between the driving field and the cavity mode with $\omega_d \neq
\omega_c$. Only in the resonance case ($\omega_d=\omega_c$), the
cavity field becomes coherent with the external driving field, as
shown by the dotted red curve in Fig.~\ref{atA}(b).

\begin{figure}
\includegraphics[width = 8.5 cm]{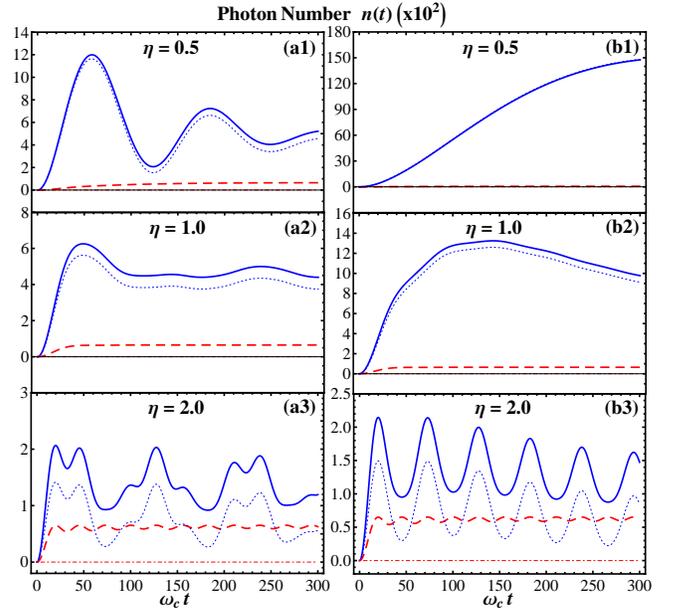}
\caption{The time evolution of the thermal fluctuation $v(t,t)$ (red
dashed and dotted-dashed curves) and the cavity intensity (blue
solid and dotted curves) at different initial temperature ($T=5$K
and $T=5$mK ) with different coupling strength. The left plots
correspond to $\omega_d = \omega_{1}$ and the right plots for
$\omega_d=\omega_{c}$.} \label{rho1}
\end{figure}

Fig.~\ref{rho1} shows the thermal fluctuation $v(t,t)$ and the
average photon number $n(t)$ in the driven cavity for different
coupling strengthes, different driving field frequencies and
different initial temperatures. According to Eq.~(\ref{solrho1emp}),
when the cavity is empty initially, the average cavity photon number
is fully determined by the driving-induced field $|y(t)|^{2}$ and
the thermal fluctuation induced correlation $v(t,t)$. As shown in
Fig.~\ref{rho1}, $v(t,t)$ almost equals to zero when $T = 5$mK. In
this case, $n(t) = |\langle a(t) \rangle|^{2} \simeq |y(t)|^{2}$,
namely, the cavity field is a pure coherent field induced by the
external driving field. When the temperature increases, the
contribution of $v(t,t)$ is increased as well. At $T=5K$, the
contribution from the thermal fluctuation is still very small in the
weak coupling, in comparison with the contribution from the
driving-induced field $y(t)$. However, $v(t,t)$ becomes comparable
with $y(t)$ in the strong coupling, as shows in
Fig.~\ref{rho1}(a3)-(b3). With further increase of the temperature,
thermal fluctuation would destroy the coherence of the cavity field
if the driving field is weak.

\begin{figure*}
\centering
\includegraphics[width = 15cm]{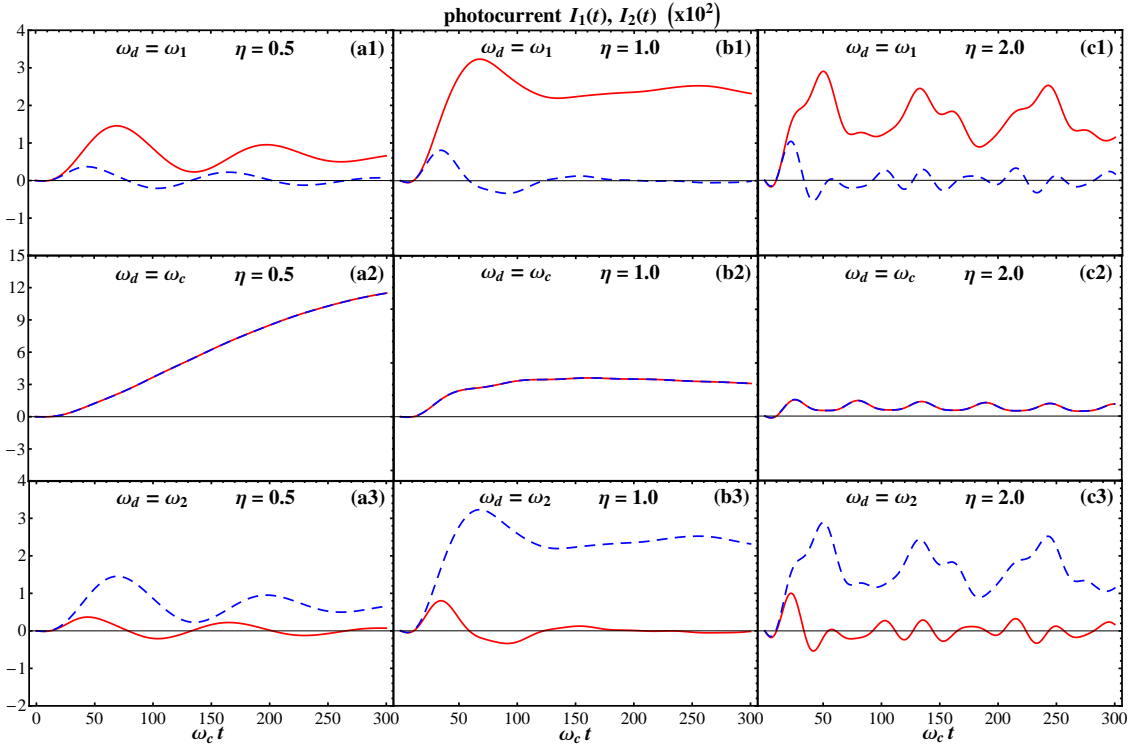}\newline
\caption{Photocurrent $I_{1}(t)$ (solid red line) and $I_{2}(t)$
(dashed blue line) in different coupling strength and driving
frequencies. Here, the initial temperature $T = 5$K.}
\label{current}
\end{figure*}

Fig.~\ref{current} shows the photocurrents $I_{1}(t)$ and $I_{2}(t)$
flowing from the cavity into the CROWs in different driving
frequencies and coupling strengthes. The frequency of the external
driving field is the crucial factor for the control of the
photocurrents. When the external driving field is in resonance to
the cavity field, the photocurrents flowing through CROW 1 and CROW
2 are equal because of the symmetric configuration, as shown in
Fig.~\ref{current}(a2)-(c2). However, when the driving field is in
resonance to the band center of CROW 1, the photocurrent $I_{1}(t)$
flowing through CROW 1 is dominated while the photocurrent
$I_{2}(t)$ flowing through CROW 2 is suppressed, see
Fig.~\ref{current}(a1)-(c1). Similarly, when the driving field is in
resonance to the band center of CROW 2, the photocurrent $I_{2}(t)$
becomes dominated and the photocurrent $I_{1}(t)$ is suppressed, as
shown by Fig.~\ref{current}(a3)-(c3). These results explicitly
demonstrate the controllability of the photonic quantum transport
through the driven cavity.

The above numerical analysis show that the photonic dynamics and
transport in an all-optical circuit can be manipulated efficiently
through the external driving signals and the internal
cavity-waveguide couplings. Photon dissipation and thermal
fluctuation appear in all the cases. In the weak coupling limit, the
cavity mode will decay so that only the driving-induced field plays
the role in the photonic transport. The strong coupling between the
cavity and the waveguides can largely suppress the cavity photon
dissipation, due to the non-Markovian memory effect. In this case,
the cavity mode can serve as a new control field in the photonic
transport. The photonic thermal fluctuation can be suppressed by
either lowing the temperature or making the device work in the high
frequency region. These general properties provide a theoretical
basis for the further development of integrated quantum photonic
circuits.

\section{SUMMARY AND PROSPECTIVE}

In summary, we have established a quantum transport theory to
describe photonic transport in photonic networks. The photonic
networks consist of all-optical circuits incorporating photonic
bandgap waveguides and driven resonators. The transient transport of
photons in photonic networks is controlled through the driven
resonators. The photonic dynamics of the driven resonators is
determined by the master equation which is derived by treating the
waveguides as an environment. The back-reactions between waveguides
and resonators and thereby the dissipation and fluctuation arisen
from the back-reactions are fully taken into account.  The
photocurrents flowing from the resonators into the waveguides that
describe the photonic transport in the network are obtained directly
from the master equation. The theory can be applied to photonic
transport in photonic networks involving many photons as well as
single photon.

Comparing with the electron transport in mesoscopic systems, we have
shown several crucial differences for the photonic transport.
Instead of the bias and gate voltage controls in electron transport,
the photonic transport is manipulated through external driving
fields applying to waveguides and resonators. Differing from the
electron transport where the bias and gate voltages controls are
imbedded into the spectral densities or the energy levels of the
system, the external driving fields applying to waveguides and
resonators turn out to be the modified driving fields acting
explicitly on the resonators, as shown in the master equation of
Eq.~(\ref{ME}). The resulting lesser Green function, see
Eq.~(\ref{lessGtaut}), and thereby the photocurrent, see
Eq.~(\ref{photocurrent1}), contain additional terms induced from the
external driving fields that explicitly determine the photonic
transport in the photonic network. In other words, the present
theory shows that the transport controls through the external fields
behave very differently for electrons and photons. Furthermore, the
explicit initial state dependence also makes the present theory
become more powerful in the study of transient transport phenomena.

As a simple illustration of the theory, we investigate the photonic
transport dynamics of the driven nanocavity coupled to two
waveguides. The cavity field, the cavity photon number as well as
the photocurrents flowing through the waveguides are solved
analytically in the weak coupling limit and are also exactly
calculated numerically for strong couplings with different driving
frequencies. Non-Markovian memory effects in the strong coupling
regime are shown explicitly. The thermal fluctuation effect to the
coherent property of the cavity field is also demonstrated.
Moreover, the analytical and numerical results show the strong
controllability of wavelength selective transport. The
controllability together with the simplicity of the photonic circuit
implementation in photonic crystals make it useful in the further
development of integrated quantum photonic circuits.

We should further point out that the present theory can also be
directly extended to investigate various photonic transport
phenomena in other electromagnetic matematerials \cite{pcmeta} and
phonon transport in heat-conducting systems \cite{heatcond}. The
generality of the present theory incorporating with the
Feynman-Vernon influence functional approach and the Keldysh's
nonequilibirum Green function technique together also makes it
powerful in the investigation of open quantum system, in particular
the nonequilibrium dynamics. The generalization of the present
theory to the nonequilibirum dynamics of ultracold atomic
Boson-Einstein condensation is in progress. Various novel devices,
such as photon entanglement through photonic crystal waveguides
\cite{Hug05,Tan11}, can be designed with the help of the present
theory. And more applications will be presented in future works.

\section*{Acknowledgement}
This work is supported by the National Science Council of ROC under
Contract No. NSC-99-2112-M-006-008-MY3. We also thank the support
from National Center for Theoretical Science of Taiwan. WMZ would
also like to thank the National University of Singapore for the warm
hospitality during his visit.

\appendix

\section{Derivation of the propagating function}

The influence functional of Eq.~(\ref{influencefunctional}) modifies
the original action of the system into an effective one, $e^{i(
S_{S}[\boldsymbol{\alpha}^{*}, \boldsymbol{\alpha}] -
S_{S}[{\boldsymbol{\alpha}'}^{*}, \boldsymbol{\alpha}'] )}
\mathscr{F}[\boldsymbol{\alpha}^{*} \boldsymbol{\alpha} ;
{\boldsymbol{\alpha}'}^{*} \boldsymbol{\alpha}']
=e^{{i\over\hbar}S_{\rm eff}[\boldsymbol{\alpha}^{*}
\boldsymbol{\alpha} ; {\boldsymbol{\alpha}'}^{*}
\boldsymbol{\alpha}'] }$ which dramatically changes the dynamics of
the driven resonators. The explicit change is manifested through the
generating function of Eq.~(\ref{propagatingfunc}) by carrying out
the path integral with respect to the effective action $S_{\rm
eff}[\boldsymbol{\alpha}^{*} \boldsymbol{\alpha} ;
{\boldsymbol{\alpha}'}^{*} \boldsymbol{\alpha}']$. While the path
integral $\mathcal{D}[\boldsymbol{\alpha}^{*} \boldsymbol{\alpha} ;
{\boldsymbol{\alpha}'}^{*} \boldsymbol{\alpha}']$ integrates over
all the forward paths $\boldsymbol{\alpha}^{*}(\tau),
\boldsymbol{\alpha}(\tau)$ and the backward paths
${\boldsymbol{\alpha}'}^{*}(\tau), \boldsymbol{\alpha}'(\tau)$ in
the Bergman complex space bounded by
$\boldsymbol{\alpha}^{*}(t)=\boldsymbol{\alpha}^{*}_{f},
\boldsymbol{\alpha}(t_0)=\boldsymbol{\alpha}_0$ and
${\boldsymbol{\alpha}'}^{*}(t_0)={\boldsymbol{\alpha}'}^{*}_0,
\boldsymbol{\alpha}'(t)=\boldsymbol{\alpha}'_{f}$, respectively.
Since $S_{\rm eff}[\boldsymbol{\alpha}^{*} \boldsymbol{\alpha} ;
{\boldsymbol{\alpha}'}^{*} \boldsymbol{\alpha}']$ is only a
quadratic function in terms of the integral variables, the path
integrals of Eq.~(\ref{propagatingfunc}) can be reduced to Gaussian
integrals so that we can use the stationary path method to exactly
carry them out \cite{Fey65}. The resulting propagating function is a
function of the stationary paths:
\begin{align}
& \mathcal{J}( \boldsymbol{\alpha}_{f}^{*} ,
\boldsymbol{\alpha}'_{f}, t | \boldsymbol{\alpha}_{i},
{\boldsymbol{\alpha}'}_{i}^{*}, t_{0} ) =  A(t) \exp\{ \frac{1}{2}[
\boldsymbol{\alpha}^{\dag}_{f}\boldsymbol{\alpha}(t) +
\boldsymbol{\alpha}^{\dag}(t_{0})\boldsymbol{\alpha}_{i} \notag \\
& + {\boldsymbol{\alpha}'}^{\dag}(t)\boldsymbol{\alpha}'_{f} +
{\boldsymbol{\alpha}'}^{\dag}_{i} \boldsymbol{\alpha}'(t_{0}) ] +
\frac{i}{2} \int_{t_{0}}^{t} [ ( {\boldsymbol{\alpha}'}^{\dag}(\tau)
- \boldsymbol{\alpha}^{\dag}(\tau) )\boldsymbol{f}(\tau)\notag \\
&  + \boldsymbol{f}^{\dag}(\tau)(
\boldsymbol{\alpha}'(\tau)-\boldsymbol{\alpha}(\tau) ) ] \} \ ,
\label{propagatingfunc1}
\end{align}
where $A(t)$ is the contribution arisen from the fluctuations around
the stationary paths and is given after
Eq.~(\ref{propagatingfunc2}). The stationary paths obey the
equations of motion :
\begin{subequations}
\label{EOM}
\begin{align}
\frac{d\boldsymbol{\alpha}(\tau)}{d\tau} & + i
\boldsymbol{\omega}(\tau)\boldsymbol{\alpha}(\tau) +
\int^\tau_{t_0}d\tau'
\mathbf{g}(\tau,\tau')\boldsymbol{\alpha}(\tau') \notag \\
+ & \int_{t_{0}}^{t}d\tau' \widetilde{\mathbf{g}}(\tau,\tau')[
\boldsymbol{\alpha}(\tau') - \boldsymbol{\alpha}'(\tau')] = - i
\boldsymbol{f}(\tau) , \\
 \frac{d\boldsymbol{\alpha}'(\tau)}{d\tau} & +
i \boldsymbol{\omega}(\tau)\boldsymbol{\alpha}'(\tau) \notag \\ - &
\int_{\tau}^{t}d\tau'
\mathbf{g}(\tau,\tau')\boldsymbol{\alpha}'(\tau')+
\int_{t_{0}}^{t}d\tau'
\mathbf{g}(\tau,\tau')\boldsymbol{\alpha}(\tau') \notag
\\ + & \int_{t_{0}}^{t}d\tau'
\widetilde{\mathbf{g}}(\tau,\tau')[\boldsymbol{\alpha}(\tau') -
\boldsymbol{\alpha}'(\tau')] = - i \boldsymbol{f}(\tau) ,
\end{align}
\end{subequations}
subjected with the boundary conditions $\boldsymbol{\alpha}(t_{0}) =
\boldsymbol{\alpha}_{i}$ and $\boldsymbol{\alpha}'(t) =
\boldsymbol{\alpha}'_{f}$. The equations of motion for
$\boldsymbol{\alpha}^{\dag}(\tau)$ and
${\boldsymbol{\alpha}'}^{\dag}(\tau)$ follow by exchanging
$\boldsymbol{\alpha}$ and $\boldsymbol{\alpha}'$ in Eq.~(\ref{EOM})
and taking a conjugate transpose, subjected to the boundary
conditions ${\boldsymbol{\alpha}'}^{\dag}(t_{0}) =
{\boldsymbol{\alpha}'}^{\dag}_{i}$ and
$\boldsymbol{\alpha}^{\dag}(t) = \boldsymbol{\alpha}^{\dag}_{f}$.

To express the master equation independent of the coherent state
representation, we shall further factorize the boundary values of
the stationary paths, $\boldsymbol{\alpha}(t_{0}) =
\boldsymbol{\alpha}_{0}$, $\boldsymbol{\alpha}'(t) =
\boldsymbol{\alpha}'_{f}$, through the following transformation:
\begin{subequations}
\label{sol}
\begin{align}
& \label{sola}  \boldsymbol{\alpha}'(\tau) -
\boldsymbol{\alpha}(\tau)
 = \bar{\boldsymbol{u}}(\tau,t)( \boldsymbol{\alpha}'_{f} -
\boldsymbol{\alpha}(t) ) , \\
& \label{solb} \boldsymbol{\alpha}(\tau) = \boldsymbol{u}(\tau,
t_{0})\boldsymbol{\alpha}_{0} + \boldsymbol{v}(\tau, t)(
\boldsymbol{\alpha}'_{f} - \boldsymbol{\alpha}(t) )  +
\boldsymbol{y}(\tau)
\end{align}
\end{subequations}
and a similar transformation for conjugate variables (with the
exchange of $\boldsymbol{\alpha}$ with $\boldsymbol{\alpha}'$ for
the boundary values $\boldsymbol{\alpha}^{\dag}(t) =
\boldsymbol{\alpha}^{\dag}_{f}$ and
${\boldsymbol{\alpha}'}^{\dag}(t_{0}) =
{\boldsymbol{\alpha}'}^{\dag}_{0}$), where $\boldsymbol{u}(\tau,
t_{0})$, $\bar{\boldsymbol{u}}(\tau,t)$, $\boldsymbol{v}(\tau, t)$
are $N$ by $N$ matrices, and $\boldsymbol{y}(\tau)$ is $1\times N$
matrix, $N$ is the dimension of the cavity system. Substituting the
solutions of Eq.~(\ref{sol}) into the equations of motion of
Eq.~(\ref{EOM}) and comparing with the boundary values, we obtain
the equations of motion for $\boldsymbol{u}(\tau, t_{0})$,
$\bar{\mathbf{u}}(\tau,t)$, $\mathbf{v}(\tau, t)$ and
$\boldsymbol{y}(\tau)$, given by Eq.~(\ref{sol2}). By multiplying
$\bar{\boldsymbol{u}}^{\dag}(\tau, t_{0})$ to Eq.~(\ref{sol2b}) and
integrating $\tau$ from $t_{0}$ to $t$, we obtain the relation
$\boldsymbol{u}(t, t_{0}) = \bar{\boldsymbol{u}}^{\dag}(t_{0}, t)$.
Similarly, we can also directly obtain the solutions of
Eqs.~(\ref{sol2c}) and (\ref{sol2d}), which are given by
Eq.~(\ref{sbuyv}).

Now, let $\tau = t_{0}$ in Eq.~(\ref{sola}) and $\tau = t$ in
Eq.~(\ref{solb}), we can express $\boldsymbol{\alpha}(t)$ and
$\boldsymbol{\alpha}'(t_{0})$ in terms of the boundary conditions
$\boldsymbol{\alpha}_{0}$ and $\boldsymbol{\alpha}'_{f}$:
\begin{subequations}
\label{sol3}
\begin{align}
\label{sol3a} \boldsymbol{\alpha}(t) = &
\boldsymbol{w}(t)\big[\boldsymbol{u}(t,t_{0})
\boldsymbol{\alpha}_{i} +  \boldsymbol{v}(t,t)
\boldsymbol{\alpha}'_{f} +  \boldsymbol{y}(t)\big] \ , \\
\label{sol3b} \boldsymbol{\alpha}'(t_{0}) = &
\boldsymbol{u}^{\dag}(t,t_{0})[ 1 - \boldsymbol{w}(t)
\boldsymbol{v}(t,t)] \boldsymbol{\alpha}'_{f} -
\boldsymbol{u}^{\dag}(t, t_{0}) \boldsymbol{w}(t) \boldsymbol{y}(t)
\notag
\\ & + [ 1 - \boldsymbol{u}^{\dag}(t, t_{0})  \boldsymbol{w}(t)
\boldsymbol{u}(t, t_{0}) ] \boldsymbol{\alpha}_{0} \ ,
\end{align}
\end{subequations}
here $\boldsymbol{w}(t) = [\boldsymbol{1} +
\boldsymbol{v}(t,t)]^{-1} = \boldsymbol{w}^{\dag}(t)$. Similarly,
$\boldsymbol{\alpha}^{\dag}(t_{0})$ and
${\boldsymbol{\alpha}'}^{\dag}(t)$ can be obtained by exchanging
$\boldsymbol{\alpha}$ and $\boldsymbol{\alpha}'$ in Eq.~(\ref{sol3})
and taking a conjugate transpose. Substituting these results with
Eq.~(\ref{sol}) into Eq.~(\ref{propagatingfunc1}), we obtain the
final form of the propagating function for the reduced density
matrix given by Eq.~(\ref{propagatingfunc2}).

\section{Analytical solution in weak coupling limit with external driving field}
In this appendix, we shall present an analytical solutions in the
weak coupling limit for the photonic network concerned in Sec.~V.
First, let us solve the photon propagating function $u(t, t_{0})$.
To do so, let $u(t, t_{0}) \equiv e^{-i\omega_c(t-t_{0})}A(t)$. Then
Eq.~(\ref{sol2b}) is reduced to
\begin{align}
\frac{dA(t)}{dt}  + \int_{0}^{t-t_{0}}dt'
\mathbf{g}(t')e^{i\omega_{c}t'}A(t-t') = 0 . \label{bma}
\end{align}
When the spectrum of the waveguides are broad enough and the
coupling between the waveguides and the system is weak, the memory
effect between the system and the reservoir can be ignored so that
the Markov limit is reached. In other words, $A(t-\tau)$ in the
integration can be replaced by $A(t)$ and the integration time $t'$
can be considered much shorter in comparison with the character time
$t$ of the cavity field \cite{An07042127}. Then,
\begin{align}
& \int_{0}^{t-t_{0}} dt'\mathbf{g}(t') e^{i\omega_{c}t'}A(t-t')
\notag
\\ & \simeq \Big[\lim_{t \rightarrow \infty
}\int_{0}^{t-t_{0}}dt'\mathbf{g}(t')e^{i\omega_{c}t'}\Big]A(t) =
(i\delta \omega_c+ \kappa) A(t) , \label{ml}
\end{align}
where $\delta \omega_c=\sum_{\alpha=1}^2 \delta \omega_\alpha$ is
 the cavity frequency shift with $\delta \omega_\alpha={\cal P}\int_{0}^{\infty}
\frac{d\omega}{2\pi} \frac{J_{\alpha}(\omega)}{ \omega_c -\omega }$.
The cavity damping rate $\kappa = \sum_{\alpha=1}^2\kappa_\alpha$
with $\kappa_\alpha= J_{\alpha}(\omega_c)/2$. Thus, the photonic
propagating function is approximated analytically by
\begin{equation}
u(t,t_{0}) \simeq e^{-(i\omega'_{c} + \kappa)(t-t_{0})} ,
\label{BMsolu}
\end{equation}
and the renormalized cavity frequency $\omega'_{c} = \omega_{c} +
\delta\omega_c$.

Next, we shall solve the correlation function $v(t,t)$ in the same
approximation. Note that $\frac{v(t,t)}{dt} \neq
\frac{v(\tau,t)}{d\tau}|_{\tau=t}$, one should not directly use
Eq.~(\ref{sol2c}) to solve $v(t,t)$. It is more convenient to take
the time derivative of the solution $v(t,t)$ given by (\ref{vtt})
and utilize the Eq.~(\ref{sol2b}), which leads to
\begin{align}
\frac{v(t,t)}{dt} + 2{\rm Re} \int_{t_{0}}^{t} d\tau &
\mathbf{g}(t-\tau) v(\tau,t) \notag \\ & = 2{\rm
Re}{\int_{t_{0}}^{t} d\tau
\tilde{\mathbf{g}}(t-\tau)\bar{u}(\tau,t)} . \label{cfl}
\end{align}
Then using the same approximation of Eq.~(\ref{ml}), we have
\begin{align}
\notag {\rm Re}\int_{t_{0}}^{t} d\tau \mathbf{g}(t-\tau) v(\tau,t) &
\simeq {\rm Re}\int_{t_{0}}^{t} d\tau g(t-\tau)
e^{i\omega_{c}(t-\tau)} v(t,t)
\notag \\ &  \simeq \kappa v(t,t) , \notag \\
{\rm Re} \int_{t_{0}}^{t} d\tau
\tilde{\mathbf{g}}(t-\tau)\bar{u}(\tau,t) & \simeq {\rm
Re}\int_{t_{0}}^{t}
d\tau \tilde{g}(t-\tau)e^{i\omega'_{c}(t-\tau)} \notag \\
& \simeq \sum_{\alpha=1}^{2}
J_{\alpha}(\omega'_c)n_{\alpha}(\omega'_{c}) ,
\end{align}
Thus, the solution of Eq.~(\ref{cfl}) is approximately given by
\begin{equation}
\label{BMsolv} v(t,t) \simeq \bar{n}(\omega'_{c},T) [1 -
e^{-2\kappa(t-t_{0})}] ,
\end{equation}
where $\bar{n}(\omega'_{c},T) = \sum_{\alpha=1}^{2}
n_{\alpha}(\omega'_{c}) J_{\alpha}(\omega'_c)/2\kappa $. The
analytical solutions of Eqs.~(\ref{BMsolu}) and (\ref{BMsolv}) are
the well-known Born-Markov limit of the cavity field coupled to a
thermal bath.

With the external driving field $f(t) =
E_{0}e^{-i\omega_{d}(t-t_{0})}$ being explicitly added, the equation
of motion for the driving-induced field $y(\tau)$ becomes:
\begin{equation}
\frac{dy(t)}{dt} + i\omega_{c}y(t) + \int_{t_{0}}^{t}d\tau
g(t-\tau)y(\tau) = -iE_{0}e^{-i\omega_{d}(t-t_{0})} .
\end{equation}
The homogeneous solution of this equation is just $u(\tau,t_{0})$
with the character frequency $\omega_c$, whose BM limit solution has
been given by Eq.~(\ref{BMsolu}). The inhomogeneous solution of
$y(t)$ must have a form $B e^{-i\omega_{d}(t-t_{0})}$, where $B$ is
a constant and $\omega_d$ is the character frequency. In the BM
limit, using the same approximation but noting the different
character frequency, we find that
\begin{align} B= \frac{E_{0}\exp{(-i\phi)}}{\sqrt{(\omega_{d} -
\tilde{\omega}_{c})^2 + \tilde{\kappa}^2}} ,
\end{align}
where $\phi=\tan^{-1}\frac{\kappa}{\omega_d-\tilde{\omega}_c}$,
$\tilde{\omega}_{c} = \omega_{c} + \sum_{\alpha=1}^{2}\delta
\tilde{\omega}_\alpha$ with the driving field induced frequency
shift $\delta \tilde{\omega}_\alpha= {\cal P}\int_{0}^{\infty}
\frac{d\omega}{2\pi} \frac{J_{\alpha}(\omega)}{ \omega_d -\omega}$
and $\tilde{\kappa} = \sum_{\alpha=1}^{2}\tilde{\kappa}_\alpha$ with
$\tilde{\kappa}_\alpha=J_{\alpha}(\omega_{d})/2$. Put the
homogeneous solution and the inhomogeneous solution together with
the initial condition $y(t_0)=0$, we obtain the driving-induced
field in the BM limit:
\begin{equation}
y(t) \simeq \frac{E_{0}\exp{(-i\phi)}}{\sqrt{(\omega_{d} -
\tilde{\omega}_{c})^2 + \tilde{\kappa}^2}} \big[
e^{-i\omega_{d}(t-t_{0})} - e^{-(i\omega'_{c} +\kappa)(t-t_{0})}
\big].
\end{equation}

From the above solution for the propagating function
$u(\tau,t_{0})$, the correlation function $v(t,t)$ and the
driving-induced field $y(t)$, it is easy to find analytically the
cavity field and the cavity photon number in the BM limit, i.e.
Eq.~(\ref{BMcavity}).
%\begin{subequations}
%\begin{align}
%\label{BMa} \langle a(t) \rangle  \simeq &
%\frac{E_{0}\exp{(-i\phi)}}{\sqrt{(\omega_{d} - \tilde{\omega}_{c})^2
%+ \tilde{\kappa}^2}} [ e^{-i\omega_{d}(t-t_{0})}\notag \\
%&~~~~~~~~~~~~~~~~~~~~~~~~~ - e^{-(i\omega'_{c}
%+\kappa)(t-t_{0})} ] \ , \\
% n(t) \simeq & \bar{n}(\omega'_{c},T)[1 - e^{-2\kappa(t-t_{0})}] \notag
%\\ &+ \frac{E_{0}^{2}}{(\omega_{d} - \tilde{\omega}_{c})^{2} +
%\tilde{\kappa}^{2}}\Big\{1 + e^{-2\kappa(t-t_{0}))} \notag \\ & ~~~
%- 2e^{-\kappa(t-t_{0})}\cos{[(\omega_{d}-\omega'_{c})(t-t_{0})]}\Big\}.
%\end{align}
%\end{subequations}
Utilizing the similar approximation in the above derivation, the
photocurrent flowing through the waveguide $\alpha=1,2$ in the BM
limit are given by Eq.~(\ref{BMsolI}).
%\begin{subequations}
%\begin{align}
% I_{1}(t) = &  -2\kappa_{1} \bar{n}(\omega'_{c})
%e^{-2\kappa(t-t_{0})} \notag \\
%& + \frac{2E_{0}^{2}\kappa_{1}}{(\tilde{\omega}_{c} -
%\omega_{d})^{2} + \tilde{\kappa}^{2}}\Big\{1+ e^{-2\kappa(t-t_{0})} \notag \\
%&  - (1 + \frac{\tilde{\kappa}_{1}}{\kappa_1})
%e^{-\kappa(t-t_{0})}\cos{[(\omega_{d} - \omega'_{c})(t-t_{0})]}
%\notag \\ &  - \frac{\delta \omega_{1} -
%\tilde{\delta}\omega_{1}}{\kappa_1}
%e^{-\kappa(t-t_{0})}\sin{[(\omega_{d} -
%\omega'_{c})(t-t_{0})]}\Big\} ,\\
%I_{2}(t) = &  -2\kappa_{2} \bar{n}(\omega'_{c})
%e^{-2\kappa(t-t_{0})} \notag \\
%& + \frac{2E_{0}^{2}\kappa_{2}}{(\tilde{\omega}_{c} -
%\omega_{d})^{2} + \tilde{\kappa}^{2}}\Big\{1+ e^{-2\kappa(t-t_{0})} \notag \\
%&  - (1 + \frac{\tilde{\kappa}_{2}}{\kappa_2})
%e^{-\kappa(t-t_{0})}\cos{[(\omega_{d} - \omega'_{c})(t-t_{0})]}
%\notag \\ &  - \frac{\delta \omega_{2} -
%\tilde{\delta}\omega_{2}}{\kappa_2}
%e^{-\kappa(t-t_{0})}\sin{[(\omega_{d} -
%\omega'_{c})(t-t_{0})]}\Big\}.
%\end{align}
%\end{subequations}
Furthermore, we can also find the source term in Eq.~(\ref{qce}) in
the BM limit by substituting the BM solution of the cavity field,
Eq.~(\ref{BMsola}), into Eq.~(\ref{source}) :
\begin{align}
S(t) = &
\frac{2\tilde{\kappa}E_{0}^{2}}{(\omega_{d}-\tilde{\omega}_{c})^{2}
+ \tilde{\kappa}^{2}} \Big\{1+
e^{-\kappa(t-t_{0})}\cos{[(\omega_{d}-\omega'_{c})(t-t_{0})]}
 \notag
\\ & + \frac{\omega_{d} - \tilde{\omega}_{c}}{\tilde{\kappa}} e^{-\kappa(t-t_{0})}
\sin{[(\omega_{d}-\omega'_{c})(t-t_{0})]}\Big\} .
\end{align}
As a self-consistent check, the BM solutions of the cavity photon
number, the driving source and the photocurrent flowing into
waveguides satisfy the quantum continuous equation:
\begin{equation}
\frac{dn(t)}{dt} = S(t) - \sum_{\alpha=1}^{2} I_{\alpha}(t) .
\end{equation}

\end{document}